\documentclass[usenatbib,twocolumn,fleqn]{mnras}
\usepackage[T1]{fontenc}
\usepackage{ae,aecompl}
\usepackage{newtxtext,newtxmath}

\usepackage[ulem={normalem,normalbf}]{changes}
\setdeletedmarkup{\ifmmode\text{\sout{\ensuremath{#1}}}\else\sout{#1}\fi}
\colorlet{Changes@Color}{red!78!black}

\usepackage{amsmath,amstext,amsopn,amssymb}
\usepackage{bm}
\usepackage[artemisia]{textgreek}
\usepackage{graphicx}

\defcitealias{2014MNRAS.437.2328B}{B14}
\defcitealias{2018ApJ...862...40C}{C18}

\title[Average Dark Matter Halo Sparsity Relations]{Average Dark Matter Halo Sparsity Relations as Consistency Check of Mass Estimates in Galaxy Cluster Samples}

\author[P.S.~Corasaniti \& Y. Rasera]
{Pier Stefano Corasaniti,$^{1,2}$\thanks{Email: \href{mailto:Pier-Stefano.Corasaniti@obspm.fr,$^{1}$}{\texttt{Pier-Stefano.Corasaniti@obspm.fr}}} and Yann Rasera$^{1}$\\
$^{1}$LUTH, UMR 8102 CNRS, Observatoire de Paris, PSL Research University,
Universit{\'e} Paris Diderot, 5 Place Jules Janssen, 92190 Meudon, France\\
$^{2}$Sorbonne Université, CNRS, UMR 7095, Institut d'Astrophysique de Paris, 98 bis bd Arago, 75014 Paris, France}

\begin{document}
\label{firstpage}
\pagerange{\pageref{firstpage}--\pageref{lastpage}}
\maketitle

\begin{abstract}
The dark matter halo sparsity provides a direct observational proxy of the halo mass profile, characterizing halos in terms of the ratio of masses within radii which enclose two different overdensities. Previous numerical simulation analyses have shown that at a given redshift the halo sparsity carries cosmological information encoded in the halo mass profile. Moreover, its ensemble averaged value can be inferred from prior knowledge of the halo mass function at the overdensities of interest. Here, we present a detailed study of the ensemble average properties of the halo sparsity. In particular, using halo catalogs from high-resolution N-body simulations, we show that its ensemble average value can be estimated from the ratio of the averages of the inverse halo masses as well as the ratio of the averages of the halo masses at the overdensity of interests. This can be relevant for galaxy clusters data analyses. As an example, we have estimated the average sparsity properties of galaxy clusters from the LoCuSS and HIFLUGCS datasets respectively. The results suggest that the expected consistency of the different average sparsity estimates can provide a test of the robustness of mass measurements in galaxy cluster samples.
\end{abstract}

\begin{keywords}
{\it cosmology}: large-scale structure of Universe; {\it galaxies}: clusters; {\it methods}: numerical
\end{keywords}

\section{Introduction}
\label{sec:intro}
There is a widespread consensus that galaxy clusters carry a wealth of cosmological information which can be retrieved from measurements of the cluster abundance, spatial clustering and mass profile \citep[see e.g.][for a review of cluster cosmology]{2011ARA&A..49..409A,2012ARA&A..50..353K}. This is because galaxy clusters are hosted in massive dark matter (DM) halos whose formation and evolution depends on the cosmic matter content, the state of cosmic expansion and the normalization amplitude of the spectrum of matter density fluctuations. Quite importantly, such measurements can provide cosmological parameter constraints complementary to those inferred from other standard cosmic probes.

Cluster cosmology is still at its infancy, but rapid progress in the field is driven by the increasing number of survey programs which target galaxy clusters through a variety of methods such us the detection of the X-ray emission of the intra-cluster gas \citep[see e.g.][]{2005ApJ...628..655V,2011A&A...534A.109P,2016A&A...592A...1P}, the imprint of the Sunyaev-Zeldovich (SZ) effect on the cosmic microwave background radiation \citep[see e.g.][]{2009ApJ...701...32S,2013ApJ...765...67M,2016A&A...594A..24P,2016A&A...594A..27P}, the richness of member galaxies in the optical and IR bands \citep[see e.g.][]{2007ApJ...660..239K,2014ApJ...785..104R} and the gravitational lensing distortion of background galaxies due to the cluster potential \citep[see e.g.][]{2011ApJ...738...41U,2012ApJS..199...25P,2012MNRAS.427.1298H}.

In the past decade cosmological constraints from galaxy cluster observations have been derived mainly from cluster abundance measurements \citep[see e.g.][]{2009ApJ...692.1060V,2016A&A...594A..24P,2017MNRAS.471.1370S}, while it has only recently been possible to derive cosmological parameter bounds from measurements of the spatial clustering of galaxy clusters thanks to the analysis of the {\it Planck} comptonization maps of the thermal SZ effect \citep{2014A&A...571A..21P,2016A&A...594A..22P}. In contrast, the use of galaxy cluster mass profile as cosmic probe remains a challenging task \citep[see e.g.][]{2010A&A...524A..68E}.

Numerical N-body simulations have shown that the density profiles of DM halos are well described by the Navarro-Frenk-White function \citep{1997ApJ...490..493N}. This characterizes the halo profile in terms of the halo mass and a fitting parameter, dubbed {\it concentration}. At a given redshift the median value of the concentration varies as function of the halo mass. Since the overall amplitude of this relation depends on the underlying cosmological model \citep[see e.g][]{2001MNRAS.321..559B,2003ApJ...597L...9Z,2004A&A...416..853D,2012MNRAS.422..185G}, it should be possible to infer cosmological parameter constraints from estimates of the concentration-mass relation. Nevertheless, several factors limit the use of the cluster concentration as cosmological proxy. For instance, astrophysical effects may alter the concentration-mass relation \citep{2010MNRAS.405.2161D,2010MNRAS.406..434M} introducing a systematic bias. Estimates of the concentration have also been shown to be sensitive to selection effects \citep{2015MNRAS.449.2024S}. Furthermore, it is far from trivial to accurately predict the median concentration-mass relation for a given cosmological model. Parametric relations have been inferred from N-body simulation studies \citep[see e.g.][]{2008MNRAS.390L..64D,2016MNRAS.457.4340K}, models based on accretion histories \citep[see e.g.][]{2003MNRAS.339...12Z,2014MNRAS.441..378L,2018MNRAS.479.1100R} and the peak height of halos \citep{2012MNRAS.423.3018P} have also been proposed in the literature to predict the cosmological dependence of the concentration-mass relation. However, all these approaches have yet to converge into a single model that can reproduce the numerical simulation results for different cosmological scenarios. No less important is that fact that these analyses should take into account the large intrinsic scatter of the concentration at any given halo mass \citep[see e.g.][]{2013ApJ...766...32B,2015ApJ...799..108D}. 

An alternative approach has been introduced by \citet[][hereafter \citetalias{2014MNRAS.437.2328B}]{2014MNRAS.437.2328B}, who have shown that the mass distribution of DM halos can be characterized in terms of the ratio of halo masses measured at radii enclosing two different overdensities. This ratio, dubbed {\it sparsity}, exhibits a number of properties that make for a reliable cosmological proxy.  First of all, at a given redshift the halo sparsity is nearly independent of the halo mass with a small intrinsic dispersion (not exceeding $\sim 20\%$ level) around a constant value that carries cosmological information encoded in the mass profile of halos. Secondly, at a given redshift the average sparsity over an ensemble of halos can be accurately predicted from prior knowledge of the halo mass function at the overdensities of interest, thus providing a quantitative framework to perform cosmological parameter inference analyses.

Recently, \citet[][hereafter \citetalias{2018ApJ...862...40C}]{2018ApJ...862...40C} have derived cosmological constraints using measurements of the redshift evolution of the sparsity of a sample of X-ray clusters. As clearly shown in \citetalias{2018ApJ...862...40C}, an additional advantage of the halo sparsity concerns its stability with respect to the impact of baryonic effects. Numerical simulation studies have shown that baryon feedback processes can have a significant impact on cluster mass estimates and cause a radial dependent mass bias \citep[see e.g.][]{2014MNRAS.442.2641V,2016ApJ...827..112B}. However, these effects appear to be significantly mitigated on halo sparsity measurements. On the one hand, sparsity measurements can target regions of the cluster profile where baryonic effects are subdominant as in regions far from the cluster core. On the other hand, being a mass ratio, the effect of systematics affecting the mass measurements results suppressed on the sparsity. This makes the halo sparsity a viable tracer of the mass distribution in clusters, a point that as been recently confirmed by the analysis of \citet{2018A&A...617A..64B}. 

An additional property of the halo sparsity concerns the fact that its average value over an ensemble of halos can be computed in distinct ways. This was already mentioned in \citetalias{2014MNRAS.437.2328B}, although never fully investigated. Here, we present a thorough study of the independent determinations of the ensemble average halo sparsity using a set of highly resolved N-body halos from the MultiDark \citep{2016MNRAS.457.4340K} and the RayGalGroupSims \citep[][]{Raygal} simulation respectively. Our results show that the coherence of the ensemble average properties of the DM halo sparsity can provide a test of the robustness of cluster sparsity measurements. To this purpose we perform a consistency check analysis of the average sparsity relations using data from the Local Cluster Substructure Survey (LoCuSS\footnote{http://www.sr.bham.ac.uk/locuss/}) catalog and HIghest X-ray FLUx Galaxy Cluster Sample (HIFLUGCS) \citep{2002ApJ...567..716R}. 

The paper is organized as follows. In Section~\ref{sec:sparsity} we introduce the halo sparsity and discuss the ensemble average sparsity relations. In Section~\ref{nbodyspar} we present the analysis of N-body halo catalogs, while in Section~\ref{locuss} and\ref{hiflugcs_spar} we describe a consistency test application to the LoCuSS and HIFLUGCS cluster dataset respectively. Finally, we present our conclusions in Section~\ref{conclu}.

\section{Halo Sparsity}\label{sec:sparsity}
The halo sparsity is defined as \citepalias[see][]{2014MNRAS.437.2328B}:
\begin{equation}
s_{\Delta_1,\Delta_2}\equiv\frac{M_{\Delta_1}}{M_{\Delta_2}},\label{spars_def}
\end{equation}
where $M_{\Delta_1}$ and $M_{\Delta_2}$ are the halo masses estimated at radii $r_{\Delta_1}$ and $r_{\Delta_2}$ enclosing the overdensities $\Delta_1$ and $\Delta_2$ with $\Delta_1<\Delta_2$. This ratio quantifies the mass excess between $r_{\Delta_1}$ and $r_{\Delta_2}$ relative to the mass enclosed in the inner radius $r_{\Delta_2}$, i.e. $s_{\Delta_1,\Delta_2}=\Delta{M}/M_{\Delta_2}+1$. Hence, the halo sparsity provides a proxy of the mass distribution inside halos without the need of specifying the functional form of the density profile. 

As shown in \citetalias{2014MNRAS.437.2328B} and \citetalias{2018ApJ...862...40C}  the general properties of the halo sparsity do not depend on the definition of the overdensity thresholds, whether expressed in units of the background density ($\rho_b$) or the critical one ($\rho_c$). \citetalias{2014MNRAS.437.2328B} have found that the largest the difference between $\Delta_1$ and $\Delta_2$ the largest the amplitude of the cosmological signal encoded in the halo sparsity. However, this does not imply that the choice of $\Delta_1$ and $\Delta_2$ can be arbitrary even with the constrain $\Delta_1<\Delta_2$. In fact, for $\Delta_1\lesssim 100$ the definition of halo as a distinct object becomes ambiguous, while for $\Delta_2\gtrsim 2000$ the sparsity probes inner regions of the halo mass profile where baryonic effect may be dominant in determining the halo mass distribution. 

It can be formally shown that halos whose density profile strictly follows the NFW profile have sparsity values which are in one-to-one correspondence with the values of the concentration parameter (see derivation in \citetalias{2014MNRAS.437.2328B}). In such a case, the sparsity would not provide additional information compared to that already encoded in the halo concentration, only a different way of retrieving it. However, not all halos are accurately described by the NFW profile. As shown in \citetalias{2014MNRAS.437.2328B}, there always exists a non-negligible fraction of halos exhibiting large deviations from the NFW function especially at large halo masses. Hence, the concentration is no longer informative of the halo mass distribution as well as its cosmological dependence. As shown in \citetalias{2014MNRAS.437.2328B} (see their Fig. 8), this is not the case of the halo sparsity, which remains close to a constant value with small intrinsic scatter even for halos with large departures from the NFW profile. Thus, it provides a probe with much wider applicability than the concentration.

Furthermore, \citetalias{2014MNRAS.437.2328B} and \citetalias{2018ApJ...862...40C} have independently shown that the average halo sparsity is a weakly dependent function of halo mass $M_{\Delta_1}$, while its ensemble average value varies with the underlying cosmological model and redshift. A direct consequence of this property is the fact that at a given redshift the average sparsity can be accurately predicted from prior knowledge of the halo mass function at $M_{\Delta_1}$ and $M_{\Delta_2}$. To show this, let us consider the equality:
\begin{equation}
\frac{dn}{d{ M}_{\Delta_2}}=\frac{dn}{d{ M}_{\Delta_1}}\frac{d{ M}_{\Delta_1}}{d{ M}_{\Delta_2}}=\frac{dn}{d{ M}_{\Delta_1}}s_{\Delta_1,\Delta_2}\frac{d\ln{{ M}_{\Delta_1}}}{d\ln{{ M}_{\Delta_2}}},\label{s_equality}
\end{equation}
where $dn/d{M}_{\Delta_1}$ and $dn/d{M}_{\Delta_2}$ are the mass functions at $\Delta_1$ and $\Delta_2$ respectively. Rearranging Eq.~(\ref{s_equality}) we can integrate over the halo ensemble mass range and compute the relation between the ensemble average of the inverse halo masses at $\Delta_1$ and $\Delta_2$. Then, assuming $s_{\Delta_1,\Delta_2}$ to be constant, the halo sparsity can be taken out of the integration giving
\begin{equation}
\int_{{M}^{\rm min}_{\Delta_2}}^{{M}^{\rm max}_{\Delta_2}}\frac{dn}{d{M}_{\Delta_2}}d\ln{{M}_{\Delta_2}}=\langle s_{\Delta_1,\Delta_2}\rangle\int_{\langle s_{\Delta_1,\Delta_2}\rangle {M}^{\rm min}_{\Delta_2}}^{\langle s_{\Delta_1,\Delta_2}\rangle {M}^{\rm max}_{\Delta_2}}  \frac{dn}{d{M}_{\Delta_1}}d\ln{{M}_{\Delta_1}},\label{sparpred}
\end{equation}
which can be solved numerically to obtain $\langle s_{\Delta_1,\Delta_2}\rangle$ for a given $dn/d{M}_{\Delta_1}$ and $dn/d{M}_{\Delta_2}$. \citetalias{2014MNRAS.437.2328B} and \citetalias{2018ApJ...862...40C} have shown that this accurately reproduces the ensemble average halo sparsity of N-body halos to better than a few percent level.

Notice that Eq.~(\ref{sparpred}) implies that the ensemble average sparsity $\langle s_{\Delta_1,\Delta_2}\rangle$ coincides with the ratio of the ensemble average of the inverse halo masses at overdensities $\Delta_1$ and $\Delta_2$ respectively:
\begin{equation}
\langle s_{\Delta_1,\Delta_2}\rangle \approx \frac{\langle 1/{M}_{\Delta_2}\rangle}{\langle 1/{M}_{\Delta_1}\rangle}.\label{sparsity_invmass}
\end{equation}

More in general we can think of $s_{\Delta_1,\Delta_2}$, $M_{\Delta_1}$ and $M_{\Delta_2}$ as random variables which taken in pairs independently probe the halo mass profile. Then, the ensemble average sparsity satisfies Eq.~(\ref{sparpred}) or equivalently Eq.~(\ref{sparsity_invmass}) provided $s_{\Delta_1,\Delta_2}$ and $1/M_{\Delta_1}$ are uncorrelated. Similarly, if also $s_{\Delta_1,\Delta_2}$ and $M_{\Delta_2}$ are uncorrelated, then
\begin{equation}\label{sparsity_averagemass}
\langle s_{\Delta_1,\Delta_2}\rangle \approx \frac{\langle {M}_{\Delta_1}\rangle}{\langle {M}_{\Delta_2}\rangle}
\end{equation}
must hold true as well. In the following, using DM halo catalogs from N-body simulations, we will test the validity Eqs.~(\ref{sparpred})-(\ref{sparsity_averagemass}), namely the fact that:
\begin{equation}
\langle s_{\Delta_1,\Delta_2}\rangle \equiv \left\langle\frac{M_{\Delta_1}}{M_{\Delta_2}}\right\rangle \approx \frac{\langle M_{\Delta_1}\rangle}{\langle M_{\Delta_2}\rangle} \approx \frac{\langle 1/{M}_{\Delta_2}\rangle}{\langle 1/{M}_{\Delta_1}\rangle} \approx \langle s^{\rm MF}_{\Delta_1,\Delta_2}\rangle,\nonumber
\end{equation} 
where $\langle s^{\rm MF}_{\Delta_1,\Delta_2}\rangle$ is the average sparsity inferred from Eq.~(\ref{sparpred}). We will refer to these different estimations as average halo sparsity relations.

Hereafter, we will specifically focus on the halo sparsity evaluated at overdensities $\Delta_1=200\rho_c$ and $\Delta_2=500\rho_c$ with $\rho_c$ being the critical density. This is motivated by the fact that we intend to develop the use of the halo sparsity as a cosmological proxy and as shown in \citetalias{2018ApJ...862...40C} at these overdensities the halo sparsity carries more pristine cosmological information encoded in halo mass profile, which is less altered by baryonic processes.

\begin{figure*}
    \includegraphics[width=1\columnwidth]{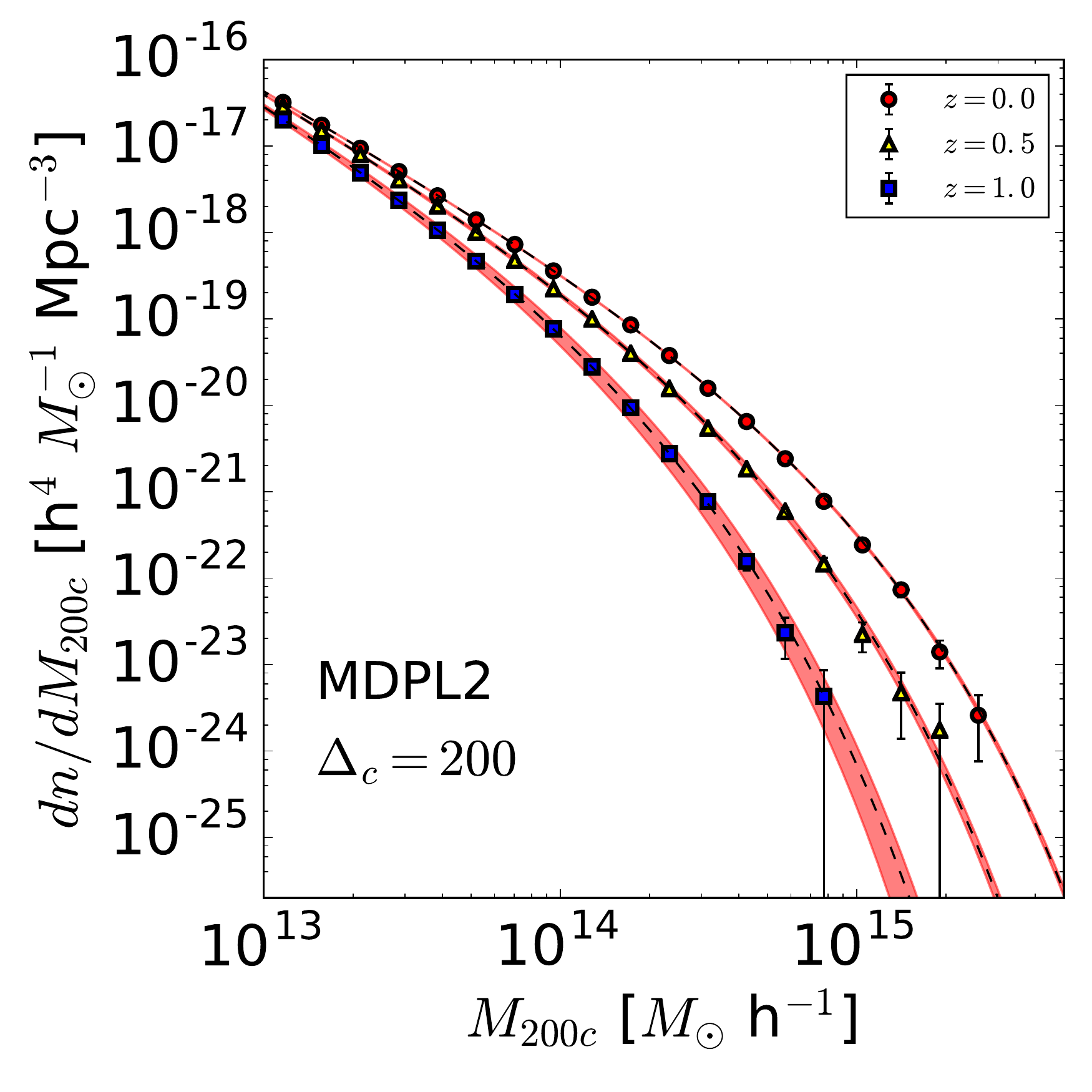}
     \includegraphics[width=1\columnwidth]{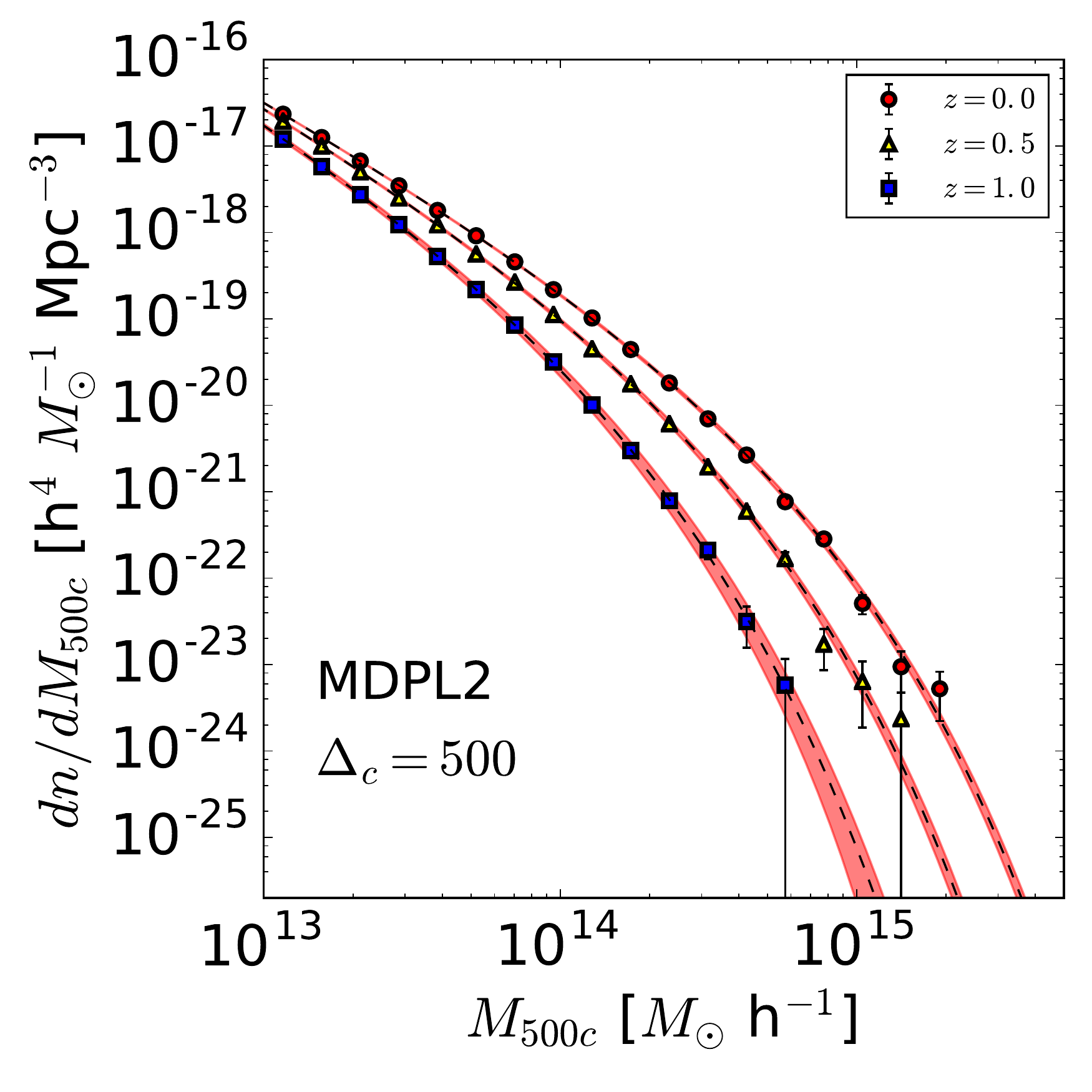}\\
     \includegraphics[width=1\columnwidth]{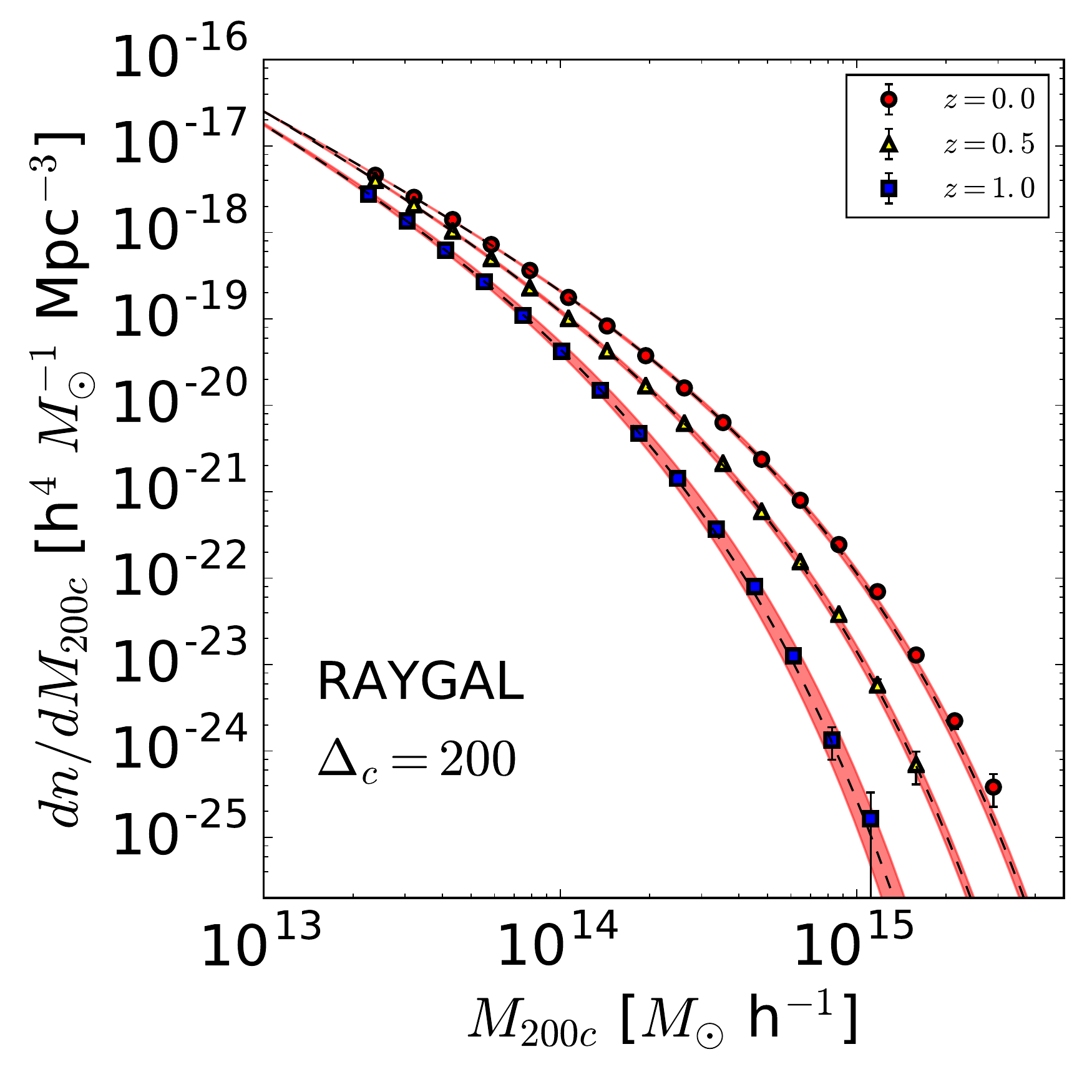}
     \includegraphics[width=1\columnwidth]{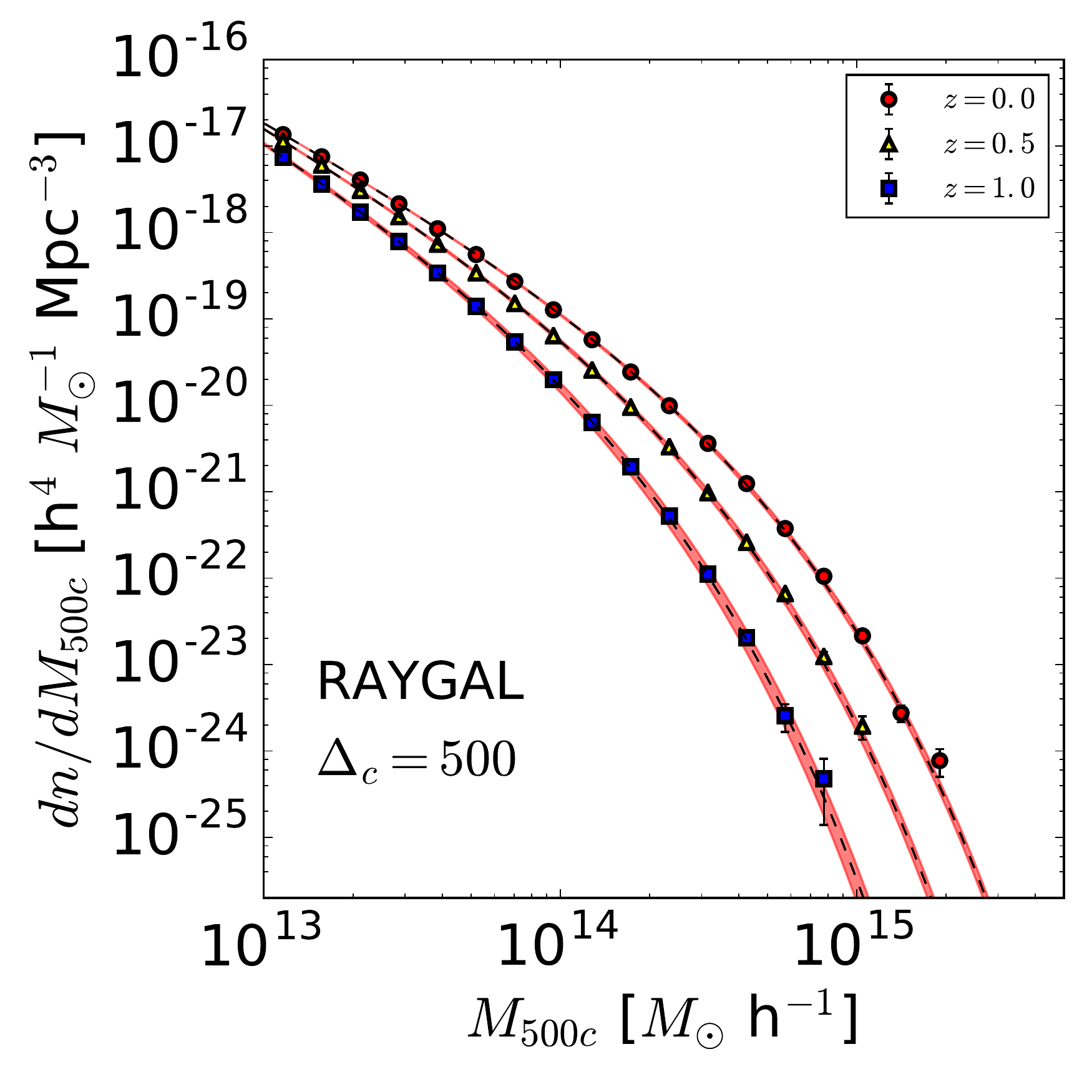}
    \caption{Halo mass function at $M_{200c}$ (left panels) and $M_{500c}$ (right panels) for $z=0,0.5$ and $1$ from the MDPL2 (top panels) and Raygal (bottom panels). The dashed lines corresponds to the best-fit ST mass functions, while the shaded areas denote the $68\%$ confidence regions.}
    \label{fig1}
\end{figure*}

\section{Average Sparsity of N-body Halos}\label{nbodyspar}

\subsection{N-body Halo Catalogs}\label{nbody}
We use N-body halo catalogs from the MultiDark-Planck2 simulation \citep{2016MNRAS.457.4340K} and the RayGalGroupSims simulation \citep[][]{Raygal} respectively. The latter has been used in \citetalias{2018ApJ...862...40C} to test the validity of Eq.~(\ref{sparpred}). Here, we extend their analysis to test the validity of the average sparsity relations.

The MultiDark-Planck2 (MDPL2) simulation \citep{2016MNRAS.457.4340K} consists of a ($1$ Gpc h$^{-1}$)$^3$ volume with $3840^3$ particles (corresponding to a mass resolution of $m_p=1.51\times 10^9$ $M_{\odot}$ h$^{-1}$) of a flat $\Lambda$CDM model with parameters calibrated against the CMB measurements from {\it Planck} \citep{2014A&A...571A..16P} with $\Omega_m=0.3071$, $\Omega_b=0.048206$, $h=0.678$, $n_s=0.96$ and $\sigma_8=0.823$. The simulation has been realized with the GADGET-2 code \citep{2005MNRAS.364.1105S}. Halos have been detected using the halo finder code Rockstar \citep{2013ApJ...762..109B}. We use the default set up with halos consisting of gravitationally bound particles only. In order to limit numerical resolution effects and given that we are interested in cluster-like halos, we only consider halos with mass $M_{200c}\ge 10^{13}$ $M_{\odot}$ h$^{-1}$ at $z=0,0.5$ and $1$ respectively. The catalogs of halo masses at overdensities $\Delta_c=200$ and $500$ (in units of the critical density) are publicly available through the CosmoSim database\footnote{https://www.cosmosim.org}. We would like to remind the reader that masses from halos with only bounded particles do not correspond to observable ones. However, in the case of isolated halos differences with respect to masses from all-particle halos are usually small (aside the case of halos affected by major mergers) and since unbounded particles will affect both masses at $\Delta_c=200$ and $500$, we expect an even smaller effect on the halo sparsity.

The RayGalGroupSims project consists of two N-body simulations of two different cosmological models realized with the RAMSES code \citep{2002AA...385..337T}. These are a standard flat $\Lambda$CDM model and a flat dark energy model with constant equation of state $w$CDM whose parameters are calibrated against the {\it WMAP} 7 yr data \citep{2009ApJS..180..330K}. These simulations have been specifically designed to generate high-resolution large volume light-cone data for ray-tracing analyses \citep[see e.g.][]{2019MNRAS.483.2671B}. The simulations cover a ($2.625$ Gpc h$^{-1}$)$^3$ volume with $4096^3$ particles (corresponding to a mass resolution of $m_p=1.88\times 10^10$ $M_{\odot}$ h$^{-1}$). The $w$CDM run has yet to be completed, hence for the purpose of this analysis we only use catalogs at $z=0,0.5$ and $1$ from the $\Lambda$CDM simulation to which we will simply refer as Raygal. The simulated $\Lambda$CDM model is characterized by the following set of cosmological parameters: $\Omega_m=0.2573$, $\Omega_b=0.04356$, $h=0.72$, $n_s=0.963$ and $\sigma_8=0.801$. We will refer to these catalogs simply as Raygal. These have been generated using the Spherical Overdensity (SO) halo detection algorithm \citep{1994MNRAS.271..676L} without unbinding. Halos are first detected using SO with the overdensity set to $\Delta_c=200$ and centered on the location of maximum density. Then, SO masses at $\Delta_c=200$ and $500$ are computed for each halo in the catalog. Consistently with the analysis presented in \citetalias{2018ApJ...862...40C} we set a conservative mass cut and consider only halos with more than $10^4$ particles. We find this to correspond to a mass limit $M_{200c}\gtrsim 2\cdot 10^{13}$ $M_{\odot}$ h$^{-1}$.

\begin{figure*}
\centering
    \includegraphics[width=0.65\columnwidth]{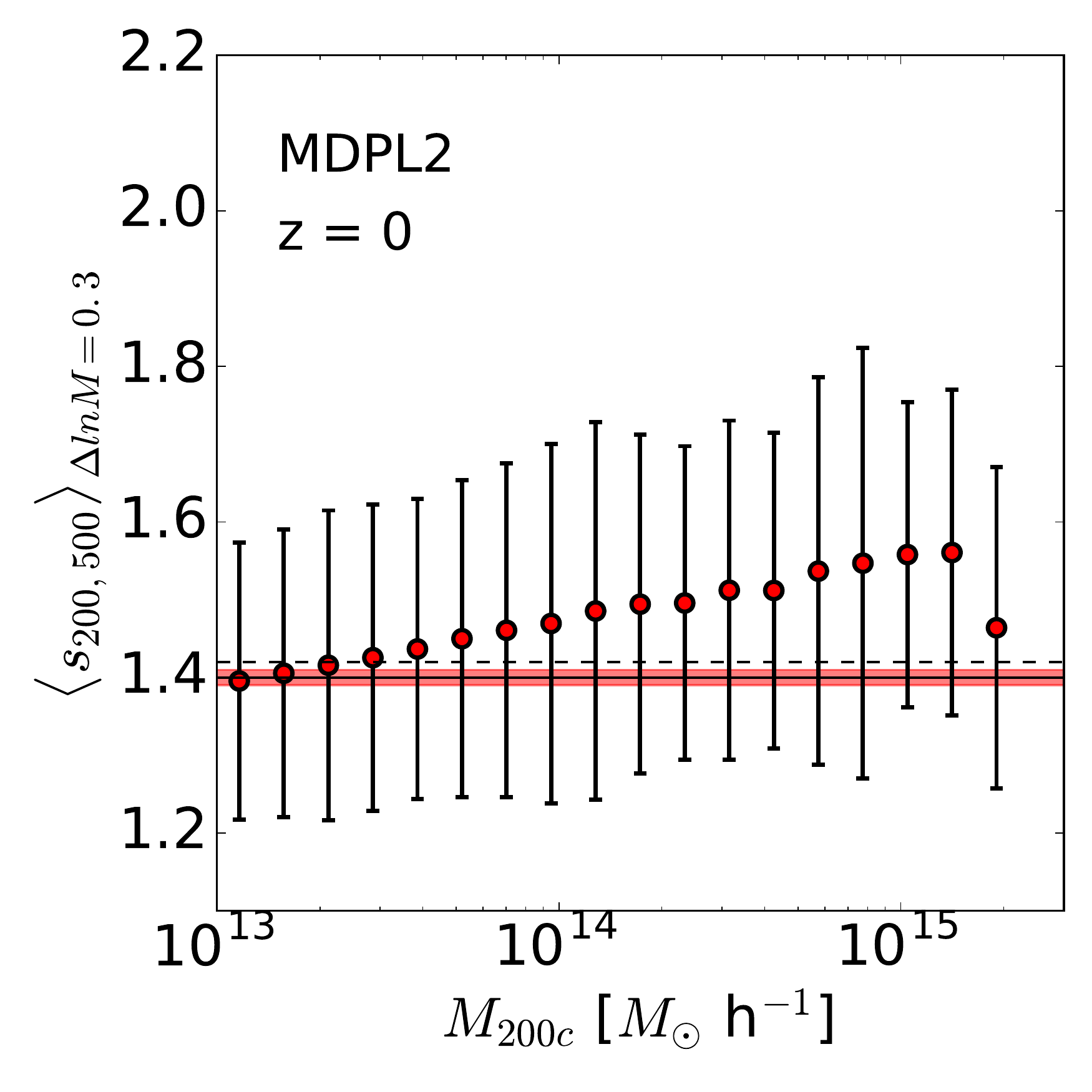}
    \includegraphics[width=0.65\columnwidth]{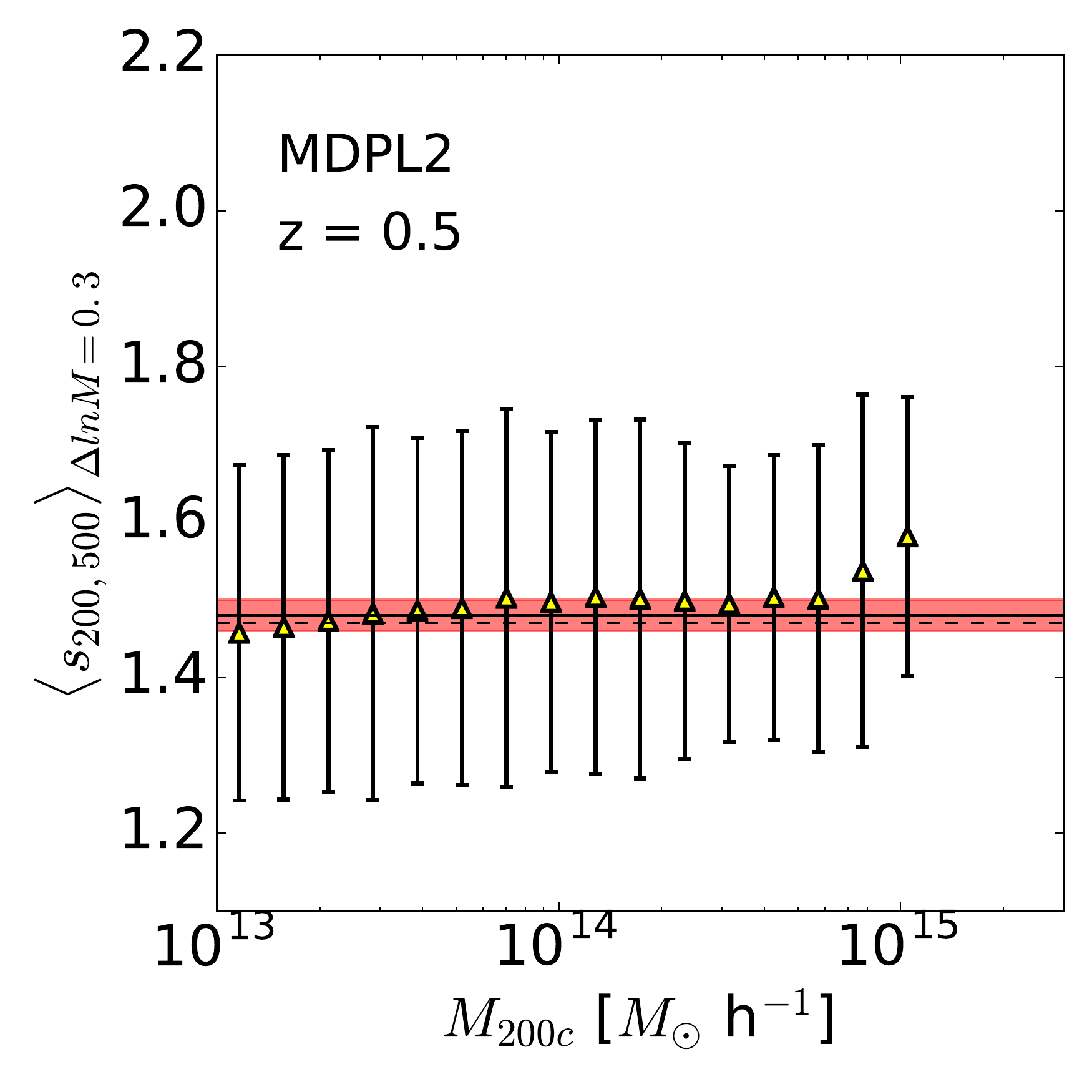}
    \includegraphics[width=0.65\columnwidth]{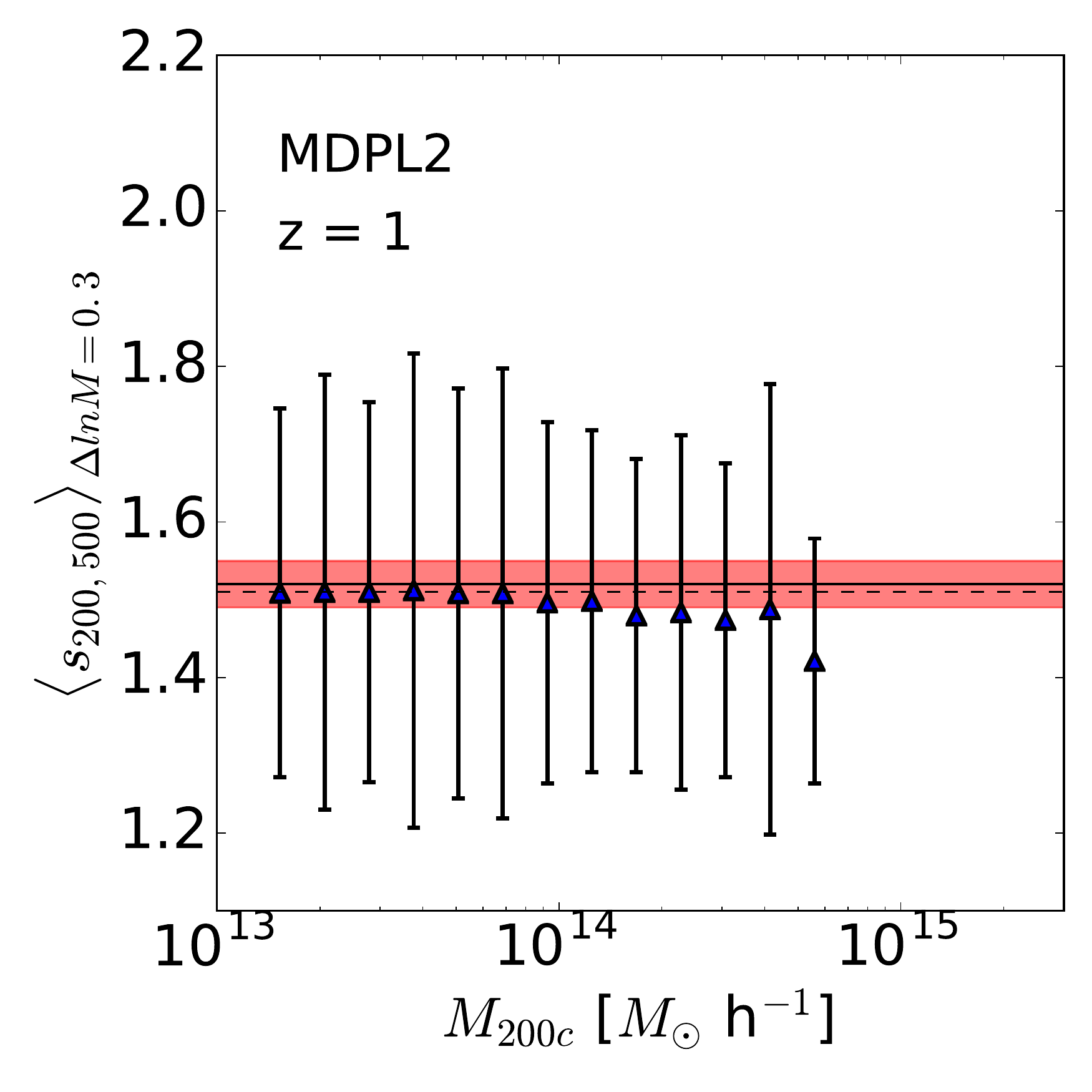}\\
    \includegraphics[width=0.65\columnwidth]{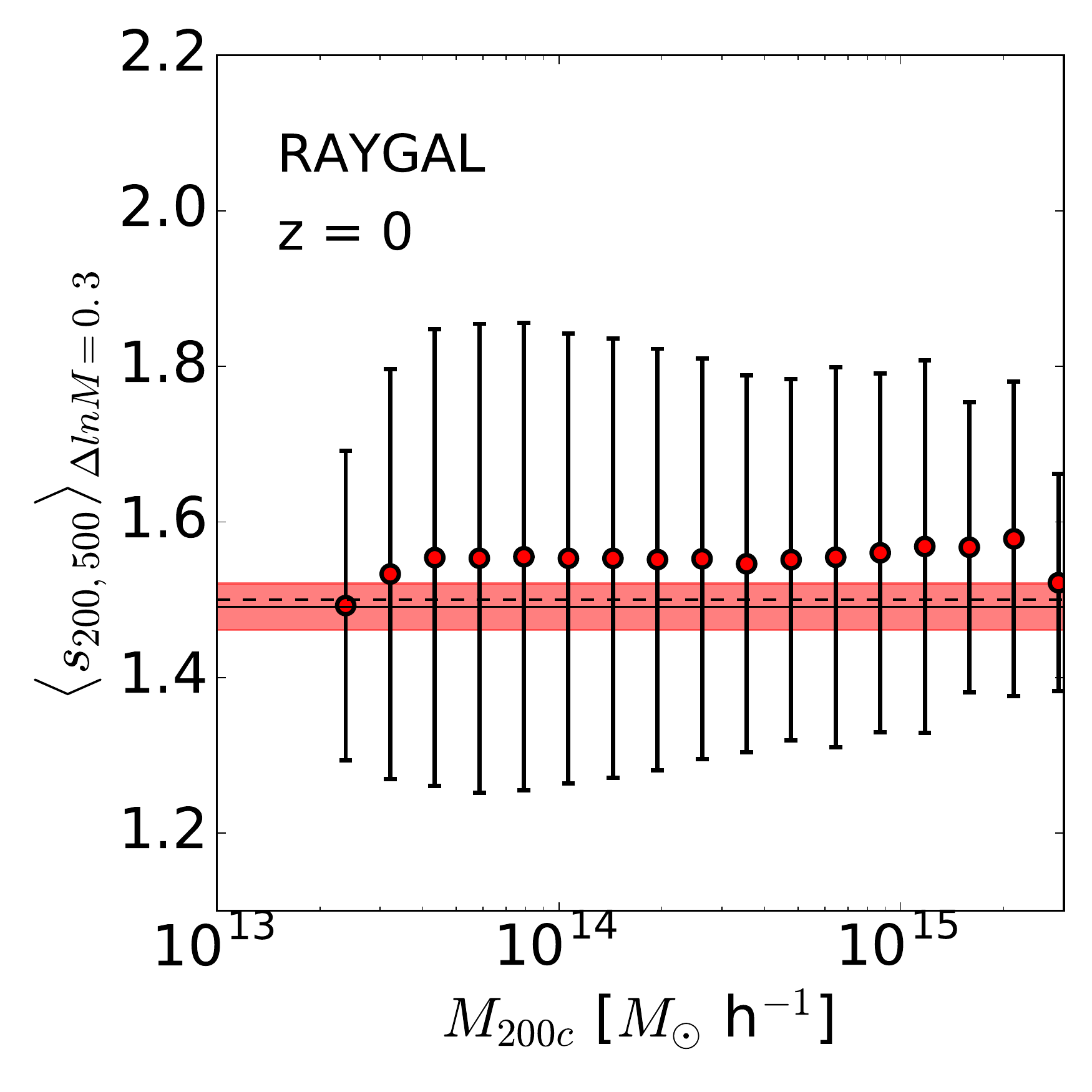}
    \includegraphics[width=0.65\columnwidth]{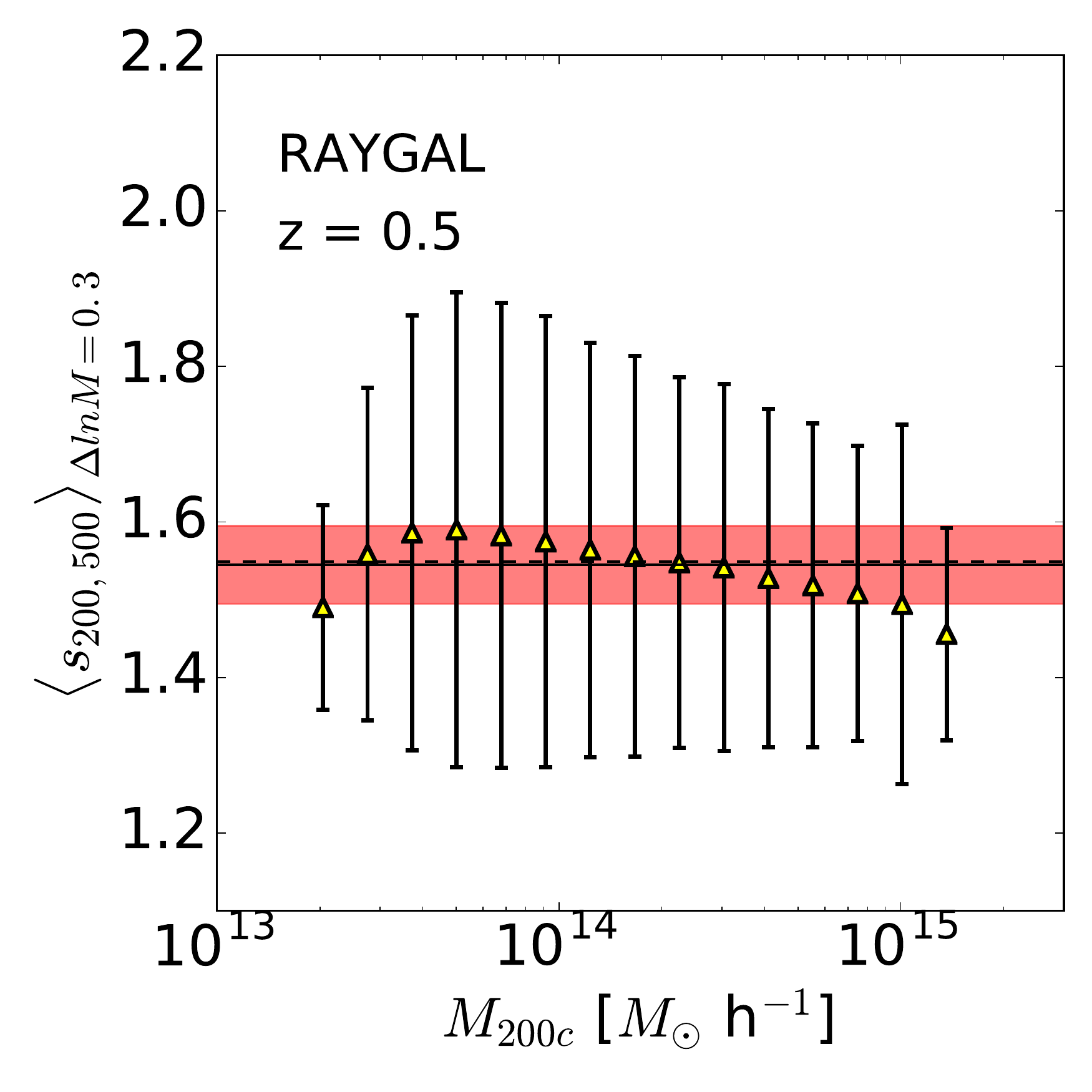}
    \includegraphics[width=0.65\columnwidth]{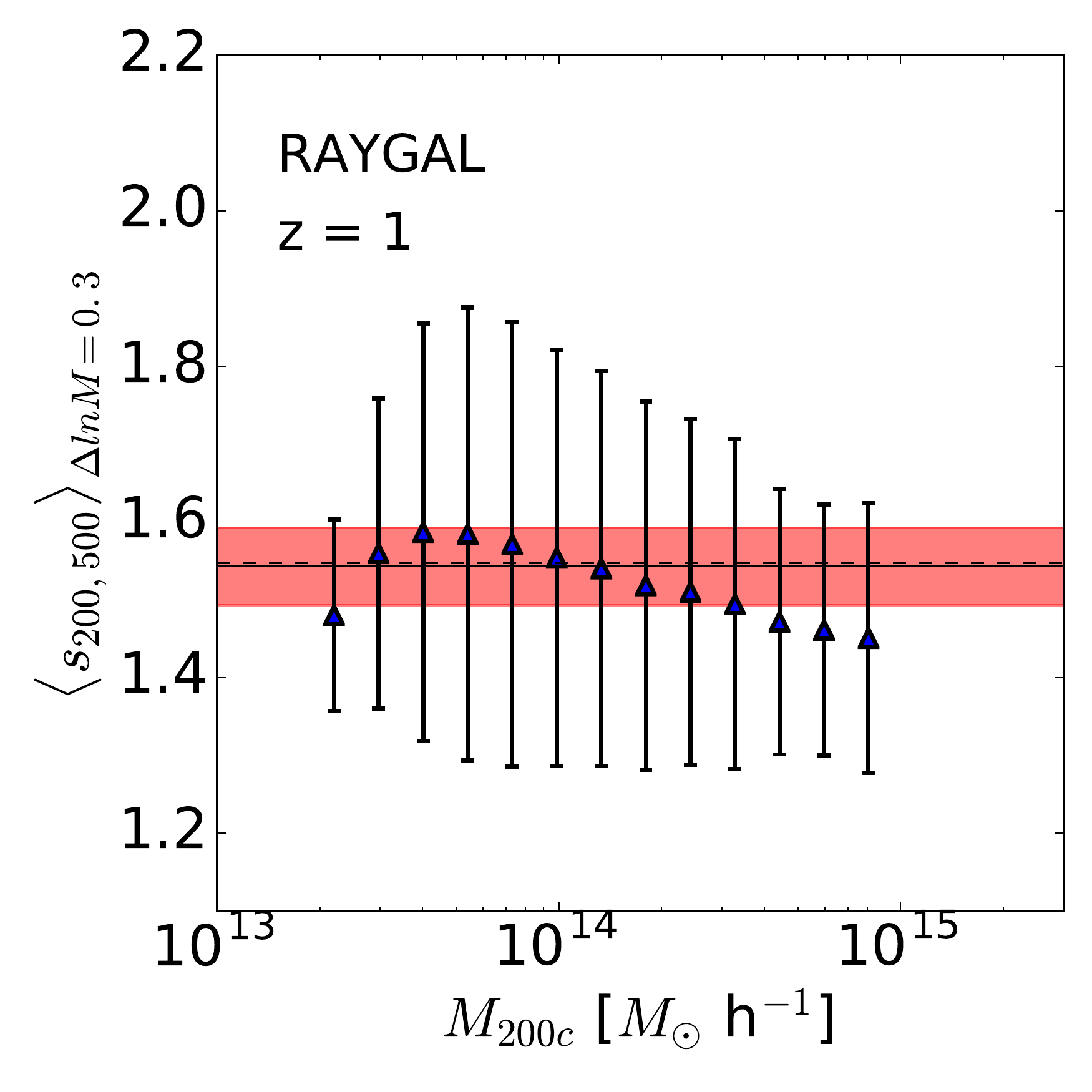}    
    \caption{Average halo sparsity as function of $M_{200c}$ in mass bins of size $\Delta \ln{M}_{200c}=0.3$ at $z=0$ (left panels) $0.5$ (central panels) and $1$ (right panels) from the MDPL2 (top panels) and Raygal (bottom panels) halo catalogs respectively. The errorbars correspond to the standard deviation around the mean value, which is dominated by the intrinsic scatter of the halo sparsity. The solid lines represent the halo ensemble average value of the sparsity as inferred from the mass function relation Eq.~(\ref{sparpred}), while the shaded regions correspond to the uncertainty due to the propagation of errors on the mass function fits. The dashed lines correspond to the ensemble average sparsity value estimated by stacking the sparsity of the halos in each catalog.}
    \label{fig2}
\end{figure*}

In Fig.~\ref{fig1} we plot the halo mass functions $z=0,0.5$ and $1$ for $M_{200c}$ (left panels) and $M_{500c}$ (right panels) respectively from the MDPL2 (top panels) and Raygal (bottom panels) halo catalogs. We fit the N-body mass function with the Sheth-Tormen (ST) formula of the multiplicity function \citep{1999MNRAS.308..119S} and determine the best-fit coefficients and the associated errors using a Levenberg-Marquardt minimisation scheme. The best-fit functions are shown in Fig.~\ref{fig1} as dashed lines, while the shaded area corresponds to the $1\sigma$ uncertainty region. We use these best-fit functions to numerically solve Eq.~(\ref{sparpred}) and predict the average halo sparsity at a given redshift for each halo catalog.

\subsection{Average Sparsity Relations}
We compute the average halo sparsity as function of $M_{200c}$ in bin size to $\Delta \ln{M}_{200c}=0.3$. This guarantees that the Poisson shot noise is negligible and contributing to the scatter around the mean sparsity value to $\lesssim 3\%$ level. We plot the results at $z=0$ (left panels), $0.5$ (central panels) and $1$ (right panels) respectively from the MDPL2 (top panels) and Raygal (bottom panels) halo catalogs in Fig.~\ref{fig2}. The errorbars represent the standard deviation which is $\approx 20\%$ level and dominated by the intrinsic scatter of the halo sparsity consistenly with the results of \citetalias{2014MNRAS.437.2328B} and \citetalias{2018ApJ...862...40C}. We find the mean sparsity to weakly vary across different mass bins with an excursion as large as $\approx 10\%$ over two decades in mass and in the redshift range considered.

In Table~\ref{tab_spars_mdpl2} we quote the values of the different average sparsity estimates at different redshifts from the MDPL2 and Raygal halo catalogs respectively. As we can see the mass dependence of the bin averaged sparsity at a given redshift only induces a percent level error on the predicted value of the average sparsity as given by Eq.~(\ref{sparpred}) compared to the ensemble average value obtained by stacking the sparsity of all halos in the catalog.

We may notice that at any given redshift the average sparsity values from the MDPL2 simulation differ from that of the Raygal simulation. This systematic difference results of the dependence of the average sparsity on the cosmological model parameters (see Section 2.3 in \citetalias{2018ApJ...862...40C}). Quite remarkably, in both cases  the different average sparsity estimates as given by Eqs.~(\ref{sparpred})-(\ref{sparsity_averagemass}) are consistent with each other to within $\sim 1-2\%$ level. This is a direct consequence of the fact that the halo sparsity $s_{200,500}$ weakly correlates with $M_{200c}^{-1}$ and $M_{500c}$ as it can be deduced by the small values of the corresponding correlation coefficients quoted in Table~\ref{tab_spars_mdpl2}. Because of this, we argue that testing the validity of Eqs.~(\ref{sparpred})-(\ref{sparsity_averagemass}) can provide a consistency test of the robustness of cluster sparsity measurements, which can then be used for cosmological parameter inference analyses of scenarios in which the structure of DM halos is similar to that found in $\Lambda$CDM N-body simulations. 

\begin{table*}
\centering
\begin{tabular}{c|c|c|c|c|c|c|c|}
\hline\hline
& $z$ & $\left\langle\frac{M_{200c}}{M_{500c}}\right\rangle$ & $\frac{\langle1/M_{500c}\rangle}{\langle1/M_{200c}\rangle}$ & $\frac{\langle M_{200c}\rangle}{\langle M_{500c}\rangle}$ & $\langle s^{\rm MF}_{200,500}\rangle\pm\sigma^{\rm MF}_{\langle s_{200,500}\rangle}$ & $r_{_{s_{200,500},M_{200c}^{-1}}}$ & $r_{s_{200,500},M_{500c}}$\\
\hline
\hline
&$0.0$ & $1.42$ & $1.41$ & $1.42$ & $1.40\pm 0.01$ & -0.12   & 0.03\\
MDPL2 &$0.5$ & $1.47$ & $1.46$ & $1.46$ & $1.48\pm 0.02$ & -0.06   &-0.05\\
&$1.0$ & $1.51$ & $1.51$ & $1.48$ & $1.52\pm 0.03$ &  0.004 & -0.12\\
\hline

&$0.0$ & $1.50$ & $1.48$ & $1.49$ & $1.49\pm 0.03$ & -0.21   & -0.03\\
Raygal &$0.5$ & $1.54$ & $1.53$ & $1.53$ & $1.54\pm 0.05$ & -0.14   & -0.08\\
&$1.0$ & $1.55$ & $1.54$ & $1.52$ & $1.54\pm 0.05$ & -0.13 &  -0.10\\
\hline
\end{tabular}
\caption{\label{tab_spars_mdpl2} Average sparsity estimates at different redshift from the MDPL2 and Raygal halo catalogs respectively. The second column gives the values of the halo ensemble average sparsity, the third and forth columns give the estimates from the ratio of the average of the inverse halo masses and ratio of the average halo masses respectively, while in the fifth column we quote the values obtained from the mass function relation Eq.~(\ref{sparpred}). The last two columns give the values of the correlation coefficients between the halo sparsity $s_{200,500}$ and $M_{200c}^{-1}$, and $M_{500c}$ respectively.}
\end{table*}

\section{Sparsity of LOCUSS Clusters}\label{locuss}
The Local Cluster Substructure Survey (LoCuSS) catalog consists of a sample of 50 clusters in the redshift interval $0.15\le z\le 0.3$ observed with the Subprime-Cam \citep{2002PASJ...54..833M} on the Subaru Telescope for which mass estimates have been derived from gravitational lensing measurements. In particular, we consider the mass measurements from \citet{2016MNRAS.461.3794O}, which have been obtained by fitting the shear profile of the observed clusters against model predictions from NFW profile assuming a power-law concentration-mass relation with amplitude and slope as free parameters. We use the mass values at $\Delta_c=200$ and $500$ from Table B1 in \citep{2016MNRAS.461.3794O} which include the corrections due to the shear calibration and the contamination of member galaxies. Then, for each cluster in the catalog we estimate the sparsity $s_{200,500}$ and evaluated the corresponding sparsity errors by propagating the cluster mass measurement uncertainties, where we have neglected the correlation between the mass estimates at the two different overdensities, since the information is not available. However, this makes our analysis only more conservative. In fact, these mass estimates are positively correlated, but neglecting positive correlations in the error propagation of the ratio of random variables results in overestimated errors. It is worth remarking that as the cluster mass measurements have been inferred assuming the NFW profile and given the relation between halo sparsity and concentration, we can only use these data to perform a consistency check of the different average halo sparsity estimations discussed in the previous sections.

\begin{figure}
    \includegraphics[width=1\columnwidth]{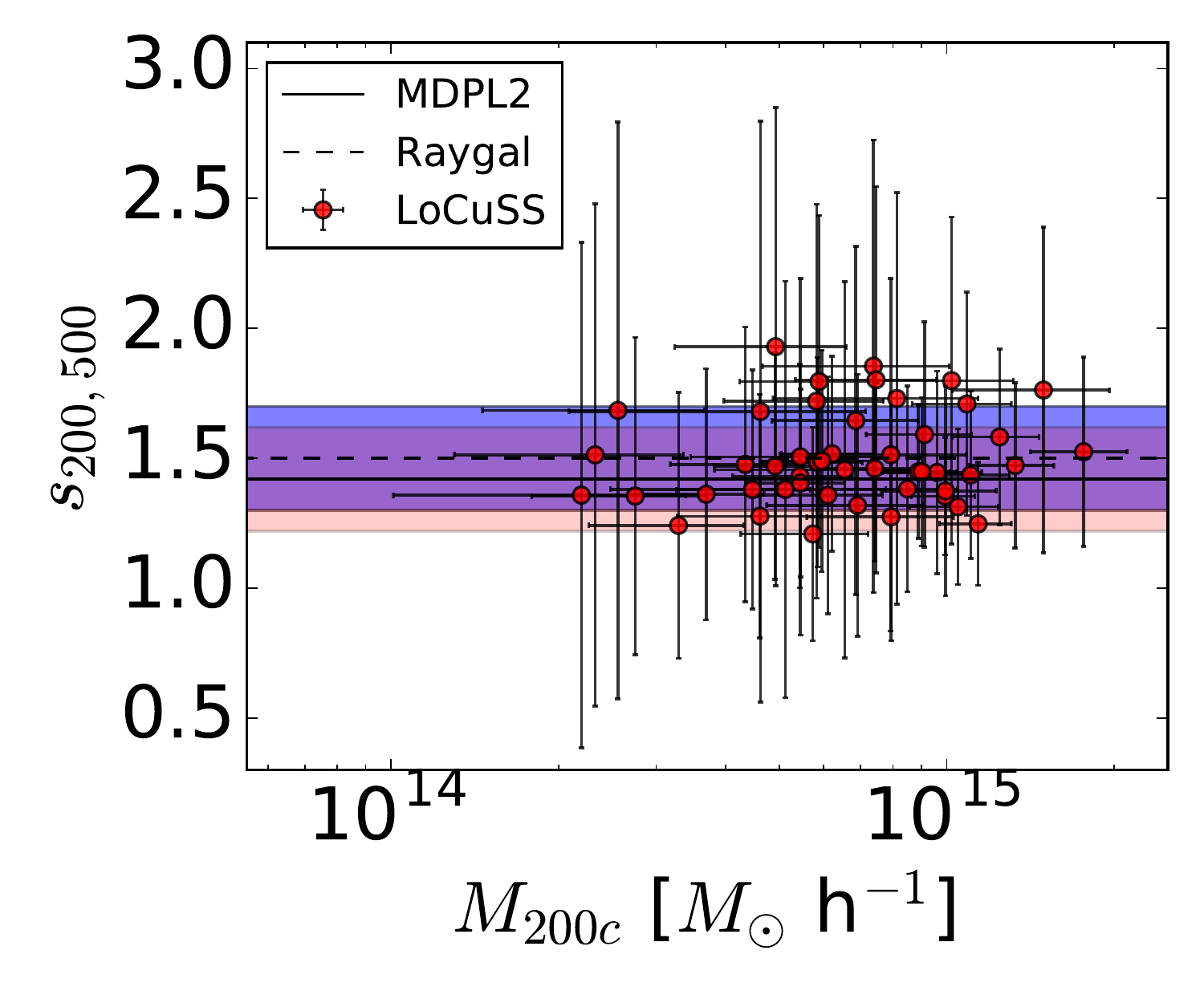}
    \caption{Sparsity $s_{200,500}$ of LoCuSS clusters. The solid (dashed) line corresponds to expected average value at $z=0$ from the MDPL2 (Raygal) simulation, while the shaded red (blue) area corresponds to the $\sim 20\%$ intrinsic scatter.}\label{fig3bis}
\end{figure}

In Fig.~\ref{fig3bis} we plot the sparsity of the LoCuSS clusters at $\Delta_c=200$ and $500$. We also show the ensemble average sparsity value at $z=0$ from the MDPL2 (solid line) and Raygal (dashed line) simulations, where the shaded area corresponds to the $\sim 20\%$ scatter. We can clearly see that cluster sparsities are largely consistent with the predictions from numerical simulations.

We find the ensemble average sparsity of the LoCuSS clusters to be:
\begin{equation}
\langle s_{200,500}\rangle \equiv \left\langle\frac{M_{200c}}{M_{500c}}\right\rangle=1.50\pm0.08,
\end{equation}
to be confronted with the value given by Eq.~(\ref{sparsity_invmass}), i.e. the ratio of the averages of the inverse halo masses:
\begin{equation}
\frac{\langle 1/M_{500c}\rangle}{\langle 1/M_{200c}\rangle}=1.49\pm 0.12,
\end{equation}
and by Eq.~(\ref{sparsity_averagemass}), i.e. the ratio of the averages of the halo masses:
\begin{equation}
\frac{\langle M_{200c}\rangle}{\langle M_{500c}\rangle}=1.49\pm 0.07.
\end{equation}
As expected all these estimates are consistent with each other well within statistical errors.
 
Given the characteristics of the LoCuSS dataset it is not possible to have a reliable estimate of the mass function at $M_{200c}$ and $M_{500c}$ to test the validity of Eq.~(\ref{sparpred}), to this purpose we will perform an analysis of a complete sample of low-redshift X-ray selected clusters which we will discuss next.

\section{Average Sparsity of HIFLUGCS Clusters}\label{hiflugcs_spar}
The HIghest X-ray FLUx Galaxy Cluster Sample \citep[HIFLUGCS][]{2002ApJ...567..716R} is a flux-limited sample of 64 galaxy clusters selected from the X-ray ROSAT All-Sky Survey \citep[][]{1999A&A...349..389V} with flux limit ${\rm F^{\rm X}_{\rm lim}}=2\times 10^{-11}$ erg s$^{-1}$ cm$^{-2}$, covering a fraction $f_{\rm sky}\approx 65\%$ of the sky. Recently, \citet{2017MNRAS.469.3738S} have reanalysed this dataset with slightly lower flux limit ${\rm F^{\rm X}_{\rm lim}}=1.7\times 10^{-11}$ erg s$^{-1}$ cm$^{-2}$ utilizing a total of 196 observations of the Chandra X-ray observatory. They have estimated hydrostatic masses of the cluster sample and measured the cluster mass function from which they have inferred cosmological parameter constraints that have been presented in \citep{2017MNRAS.471.1370S}. Since the majority of HIFLUGCS clusters are at redshift $z<0.1$, this dataset provides a homogeneous sample of cluster mass estimates useful for testing the average halo sparsity consistency relations discussed in Section~\ref{sec:sparsity}. 

\subsection{HIFLUGCS Hydrostatic Masses \& Baryon Corrections}
Hydrostatic mass estimates of the HIFLUGCS catalog have been obtained by \citet{2017MNRAS.469.3738S}. The X-ray measurements of the cluster temperature profiles are limited to a maximum extraction region of $\approx 12$ arcmin or equivalently $670$ kpc, which is smaller than $r_{500c}$ (i.e. the radius enclosing an overdensity $500\rho_c$). Consequently, the inferred cluster masses depend on the extrapolation scheme that has been adopted to solve the hydrostatic equilibrium equations.

Here, we consider the {\it kT extrapolate} (kT) and {\it NFW-Freeze} (NFW-F) schemes for which \citet{2017MNRAS.469.3738S} have derived  mass estimates at  $\Delta_c=200$ and $500$ as quoted in their Table B3. The former method is a simple extrapolation based on parametrized analytical models of the temperature and surface brightness. The latter approach assumes a NFW-profile for the DM component in combination with the empirical concentration-mass relation from \citet{2013ApJ...766...32B}. Because of this we may expect that NFW-F based sparsity estimates satisfy the ensemble average sparsity relations. This is not necessarily the case for the sparsity values obtained from the kT-masses for which no assumption on the cluster mass profile has been made. Hence, the validity of the sparsity relation can be a consistency test of the reliability of such mass measurements. Nevertheless, all these mass estimates are the result of an extrapolation and as such unaccounted systematics may manifest in departures from sparsity ensemble average properties which we have derived from the analysis of N-body simulations.

Numerical studies have shown that feedback of Active Galactic Nuclei (AGN) on the baryon distribution can alter the DM halo mass at a given overdensity with respect to that expected from DM-only simulations \citep[see e.g.][]{2014MNRAS.442.2641V,2016ApJ...827..112B}. This induces a radial dependent mass bias and cause a systematic error on the determination of the halo sparsity. As shown in \citetalias{2018ApJ...862...40C} the amplitude of this effect strongly depends on the target overdensity as well as the overall DM halo mass. In the case of $s_{200,500}$ the radial dependent mass bias results into a systematic shift of the ensemble average value of order $\lesssim 4\%$ with respect to the DM-only case. The effect is largest for $M_{200c}\sim 10^{13}$ M$_{\odot}$ h$^{-1}$ and reduces to sub-percent level at $M_{200c}\gtrsim 10^{14}$ M$_{\odot}$ h$^{-1}$.

In order to assess the impact of baryonic processes on the hydrostatic mass estimates we correct the HIFLUGCS masses for the baryon radial dependent mass bias using the fitting formula provided in \citep{2014MNRAS.442.2641V}:
\begin{equation}
\log_{10}\left(\frac{{M}_{\Delta}}{{M}_{\Delta}^{\rm c}}\right)=A_{\Delta}+\frac{B_{\Delta}}{1+\exp{\left(-\frac{\log_{10}{M}_{\Delta}^{\rm c}+C_{\Delta}}{D_{\Delta}}\right)}}, \label{barcorr}
\end{equation}
where the value of the coefficients $A_{\Delta}$, $B_{\Delta}$, $C_{\Delta}$ and $D_{\Delta}$ depend on the simulated baryon feedback model. For each cluster with hydrostatic mass $M_{\Delta}$ we determine the baryon corrected DM-only mass $M_{\Delta}^{\rm c}$ by numerically inverting Eq.~(\ref{barcorr}). We set the coefficients in Eq.~(\ref{barcorr}) to the AGN-8.0 model values quoted in table 2 of \citep{2014MNRAS.442.2641V}, since, as pointed out by the authors, the AGN-8.0 model simulation reproduces the observed X-ray cluster profiles \citep{2014MNRAS.441.1270L}. 

\begin{figure}
    \includegraphics[width=1\columnwidth]{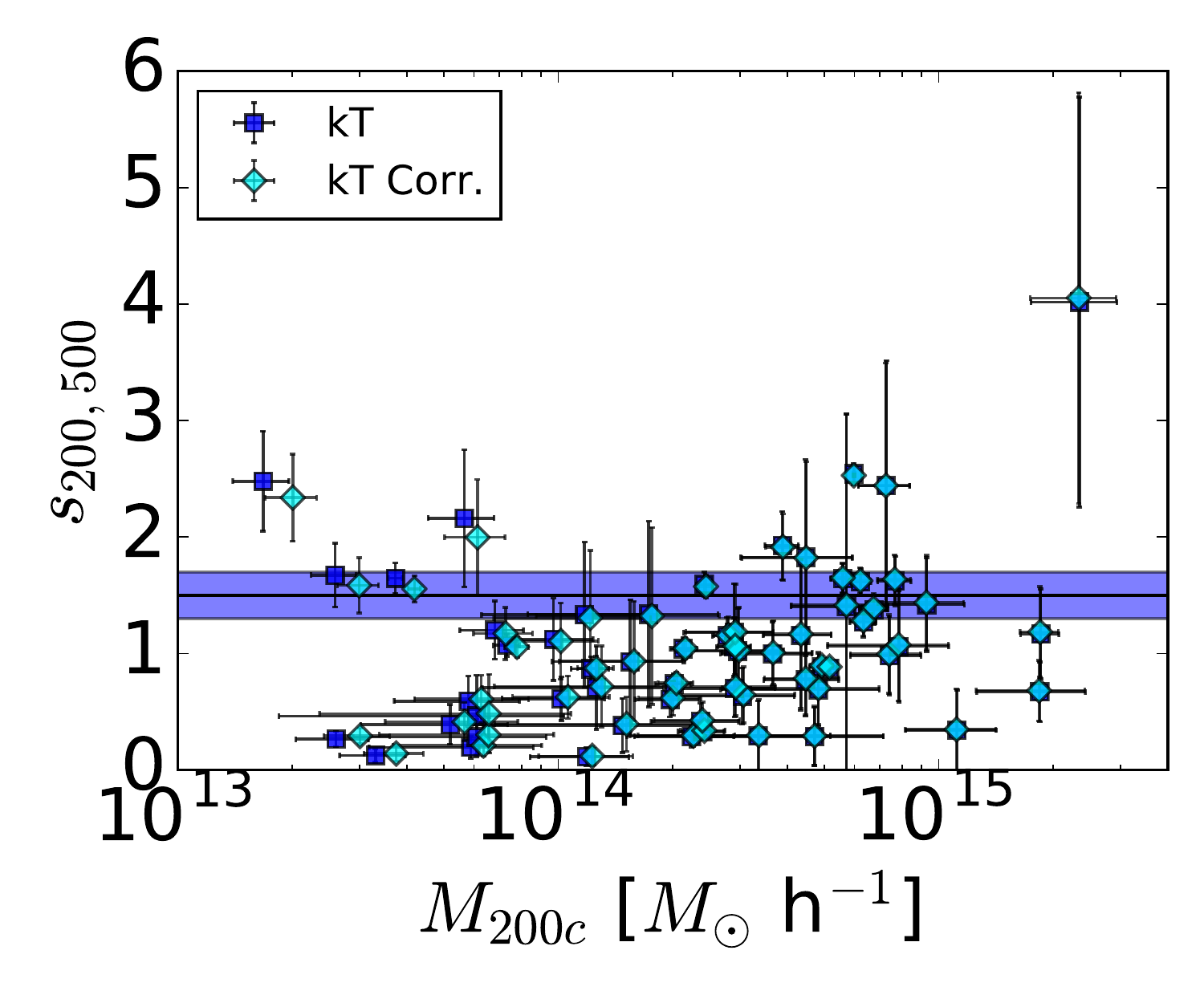}\\
    \includegraphics[width=1\columnwidth]{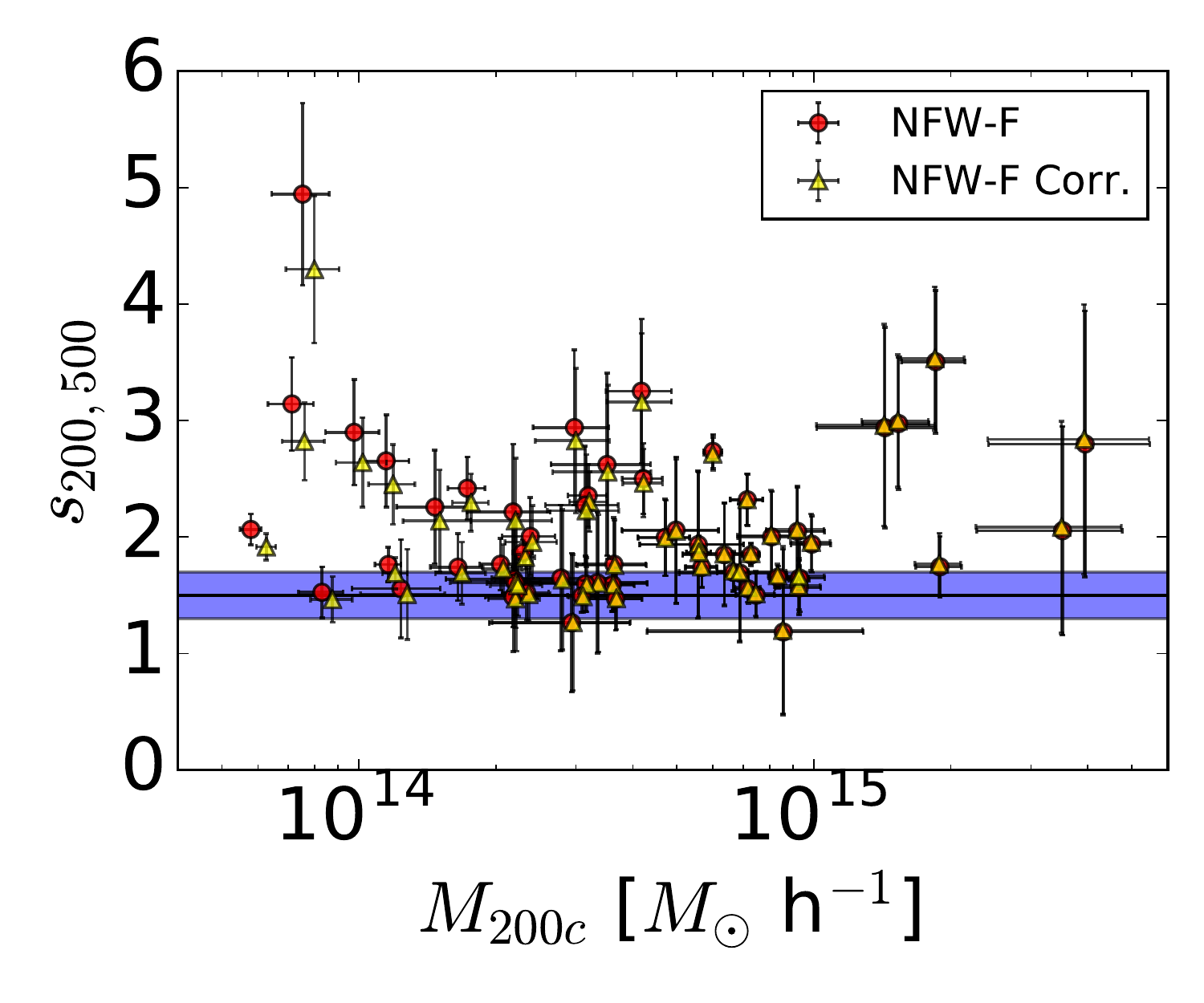}
    \caption{Sparsity $s_{200,500}$ of HIFLUGCS clusters at $z<0.1$. Top panel: sparsity from the kT-mass estimates without (blue squares) and with (cyan diamonds) baryon corrections. Bottom panel: sparsity from NFW-F mass estimates without (red circles) and with baryon corrections (yellow triangles). The solid line corresponds to the expected average value at $z=0$ from the Raygal simulation, while the shaded blue area correponds to the $\sim 20\%$ intrinsic scatter.}
    \label{fig3}
\end{figure}

In Fig.~\ref{fig3}, we plot the sparsity of the HIFLUGCS catalog for clusters at $z<0.1$ from the kT (top panel) and NFW-F (bottom panel) mass estimates without and with the baryon correction. Again we have estimated the errors on the cluster sparsity by propagating the mass uncertainties, while neglecting the correlation between the mass measurements at the two different overdensities (such information not being available). However, as already mentioned in Section~\ref{locuss}, this only results in overestimated errors and thus a more conservative analysis of the HIFLUGCS dataset.

We may notice from Fig.~\ref{fig3} that the baryon corrected sparsities, independently of the mass extrapolation scheme, remain well within the errors of the uncorrected mass measurements. The baryon correction becomes negligible for $M_{200c}\gtrsim 4\cdot 10^{14}$ $M_{\odot}$ h$^{-1}$, which is expected for very massive systems. In Fig.~\ref{fig3} we also plot the expected ensemble average sparsity from the Raygal simulation at $z=0$ (solid line) with a $20\%$ dispersion (blue shaded area). We can see that there is a substantial differences between the kT sparsities and those obtained from the NFW-F masses. 

The values inferred from the kT scheme are prevalently smaller than the Raygal expectation (about $1/3$ of the sample) and smaller than unity. This is quite surprising since by definition $s_{200,500}=\Delta{M}/M_{500c}+1>1$. This is due to the fact that for these clusters $M_{500c}^{\rm kT}>M_{200c}^{\rm kT}$. This puzzling outcome is an artefact of the kT-extrapolation of the cluster masses at radii well beyond the radial range covered by the observations. In fact, as already pointed out by \citet{2017MNRAS.469.3738S}, the kT masses are inferred by solving the hydrostatic equilibrium equations using only parametrized functions of the temperature and surface brightness profile, which are fitted against the data. Consequently, uncertainties are largely underestimated beyond the range covered by the data, where these parametrized profiles (which are not physically motivated) may significantly deviate from that of the cluster external regions. This can be clearly seen in the case of A1650 \citep[see Figure C16 in][]{2017MNRAS.469.3738S} for which the inferred kT mass profile shows an unphysical cut-off beyond $r_{500c}$. 
 
In the case of NFW-F masses about $\sim 2/3$ of the clusters are within $1\sigma$ of the simulation result with sparsity values consistently larger than unity. Nonetheless, we find a large scatter toward larger sparsity values. Again, we believe this to be a consequence of the extrapolation of the cluster masses at radii much larger than those probed by the X-ray observations (even when a NFW model with a specific concentration-mass relation is adopted). As an example, NGC4636 is the cluster with the largest NFW-F sparsity with the baryon corrected value to $s_{200,500}^{\rm NFW-F\,Corr.}=4.30\pm 0.63$. We find this to be more than $5\sigma$ away from that expected from the Raygal simulation. However, this is also the cluster in the HIFLUGCS catalog with measurements of the temperature profile covering the smallest radial interval, $\lesssim 50$ kpc \citep[see Fig. C59 in][]{2017MNRAS.469.3738S}.

\subsection{HIFLUGCS Luminosity-Mass Relation \& Mass Function}
In order to test the validity of Eq.~(\ref{sparpred}) we need to derive an analytical fit of the HIFLUGCS halo mass function at $\Delta_c=200$ and $500$. However, as the HIFLUGCS dataset is a flux limited sample of X-ray clusters, the evaluation of the mass function requires several intermediate steps:
\begin{itemize}
\item[-] we fit the HIFLUGCS luminosities and cluster masses to derive the luminosity-mass relation for each mass definition and extrapolation scheme (with and without baryon corrections);
\item[-] we compute the survey volume for each of the inferred luminosity-mass relations and evaluate the corresponding mass functions;
\item[-] we perform a $\chi^2$-analysis of the mass function estimates to determine the best-fit parameters and the associated errors of parametrized analytical fitting function;
\item[-] we use the halo mass function fits to numerically solve Eq.~(\ref{sparpred}) for each extrapolation scheme (with and without baryon corrections).
\end{itemize}

Formally, the halo mass function at a given overdensity $\Delta$ reads as:
\begin{equation}\label{massfunction_def}
\frac{dn}{d\ln{M}_{\Delta}} \equiv \frac{N_{\rm cl}({M}_{\Delta})}{V_{\rm eff}({M}_{\Delta})\Delta\ln{M_{\Delta}}},
\end{equation}
where $N_{\rm cl}({M_{\Delta}})$ is the number of clusters in mass bins of size $\Delta{\ln{M_{\Delta}}}$ and $V_{\rm eff}({M}_{\Delta})$ is the effective survey volume. For X-ray flux limited sample, the effective survey volume reads as \citep[see e.g.][]{2009ApJ...692.1033V}:
\begin{equation}\label{vm_eff}
V_{\rm eff}({M}_{\Delta})=\int_{z_{\rm min}}^{z_{\rm max}}dz\, \frac{dV}{dz}(z)A(f_{\rm X},z)\int_{L_{{\rm X},{\rm min}}(z)} dL_{\rm X}\,p(L_{\rm X} \vert {M}_{\Delta}),
\end{equation}
where $dV/dz$ is the cosmological volume factor \citep[for a formal definition see e.g.][]{1999astro.ph..5116H}, $A(f_{\rm X},z)$ is the effective survey area, $L_{{\rm X},{\rm min}}(z)=4\pi d_L^2(z) {\rm F}^{\rm X}_{\rm lim}$ with $d_L$ being the luminosity distance and 
\begin{equation}\label{pdflm}
p(L_{\rm X}\vert {M}_{\Delta}) =\frac{1}{\sqrt{2\pi\sigma^2}}\exp\left[-\frac{(\ln{L}-\ln{\bar{L}_{\rm X}({M}_{\Delta})})^2}{2\sigma^2}\right],
\end{equation}
where $\bar{L}_{\rm X}({M}_{\Delta})$ is the average X-ray luminosity-mass relation. 

\begin{figure}
    \includegraphics[width=1.\columnwidth]{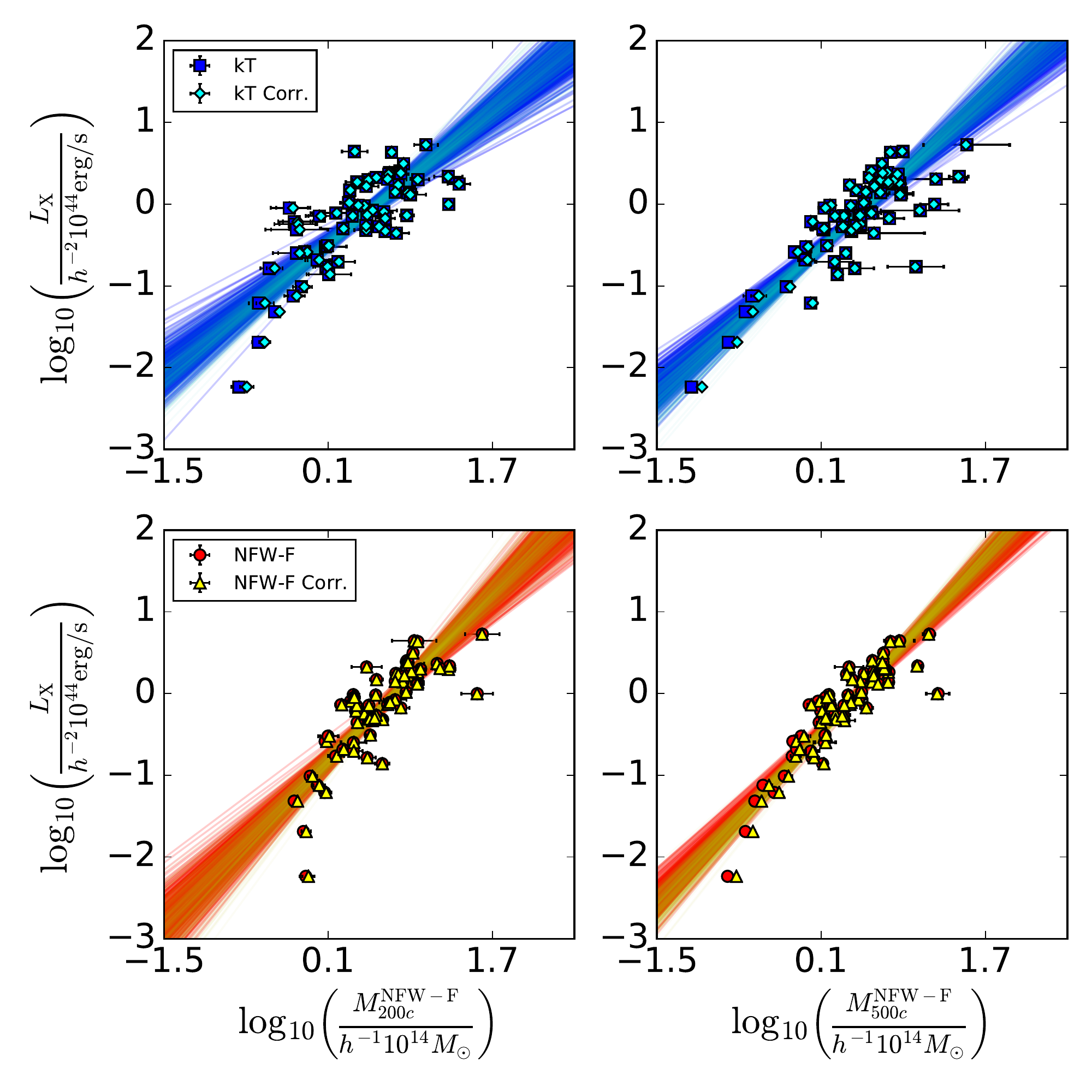}
    \caption{$L_{x}-{M}_{\Delta}$ relation for $\Delta_c=200$ (left panels) and $500$ (right panels) for the kT (left panels) and NFW-F (right panels) mass extrapolation schemes respectively. The shaded area corresponds to the $68\%$ confidence region (not including the intrinsic scatter $\sigma^2$).}
    \label{fig4}
\end{figure}

In order to evaluate $\bar{L}_{\rm X}({M}_{\Delta})$ we follow \citet{2017MNRAS.471.1370S}, who have assumed a log-linear relation:
\begin{equation}\label{linear_LM}
\log_{10}\left(\frac{\bar{L}_{\rm X}}{{\rm h}^{-2} 10^{44}{\rm erg\,s^{-1}}}\right)=A_{\rm LM}+B_{\rm LM}\cdot\log_{10}\left(\frac{M_{\Delta}}{h^{-1} 10^{14}{M}_{\odot}}\right)
\end{equation}
where $A_{\rm LM}$ and $B_{\rm LM}$ are fitting parameters\footnote{In \citet{2017MNRAS.471.1370S} the mass $M_{\Delta}$ in Eq.~(\ref{linear_LM}) is normalized to $10^{15}M_{\odot}{\rm h}^{-1}$.}. These have been estimated in \citet{2017MNRAS.471.1370S} by performing a Markov Chain Monte Carlo cosmological model parameter analysis of the HIFLUGCS cluster mass function using the NFW-F mass estimates at $M_{500c}$. As we need to derive Eq.~(\ref{linear_LM}) for the different mass definition and extrapolation scheme, we perform an independent statistical analysis to constrain $A_{\rm LM}$, $B_{\rm LM}$ as well as the intrinsic variance $\sigma^2$ entering in Eq.~(\ref{pdflm}). To this purpose we use the \textsc{\mbox{linmix}}\footnote{\url{http://linmix.readthedocs.io/en/latest/}} package, which implements a Bayesian hierarchical model described in \citep{2007ApJ...665.1489K}. The mean and standard deviation of $A_{\rm LM}$, $B_{\rm LM}$ and $\sigma^2$ are quoted in Table~\ref{tab2}, while in Fig.~\ref{fig4} we plot the inferred $\bar{L}_{\rm X}-{M}_{\Delta}$ relations against the HIFLUGCS dataset for the kT (top panels) and NFW-F extrapolation (bottom panels) schemes. Here, it is worth to comparing our results for the NFW-F masses at $M_{500c}$ with those quoted in Table 1 of \citep{2017MNRAS.471.1370S} for the case "Bayesian", which have been derived with a code implementing the hierarchical model presented in \citep{2007ApJ...665.1489K}. Because of the different mass normalization we have adopted in Eq.~(\ref{linear_LM}), the constraints on $A_{\rm LM}$ differ from those found in \citep{2017MNRAS.471.1370S}. In contrast, we find exactly the same value for the slope $B_{\rm LM}$ and the intrinsic scatter $\sigma_2$.

\begin{table*}
\centering
\begin{tabular}{c|c|c|c|c}
\hline
 & kT & kT Corr. & NFW-F & NFW-F Corr.  \\
\hline
$\Delta_c=200$ & & & &\\
\hline
A$_{\rm LM}$ & $-0.56\pm 0.06$ & $-0.58\pm 0.07$ & $-0.94\pm 0.08$ & $-0.95\pm 0.08$ \\
B$_{\rm LM}$ & $1.09\pm 0.09$ & $1.00\pm 0.11$ & $1.25\pm 0.12$ & $1.27\pm 0.12$ \\
$\sigma^2$ & $0.09\pm 0.02$ & $0.13\pm 0.03$ & $0.11\pm 0.23$ &$0.11\pm 0.02$ \\
\hline
$\Delta_c=500$ & & & &\\
\hline
A$_{\rm LM}$ & $-0.62\pm 0.05$ & $-0.64\pm 0.05$ & $-0.57\pm 0.04$ & $-0.61\pm 0.04$ \\
B$_{\rm LM}$ & $1.09\pm 0.09$ & $1.14\pm 0.10$ & $1.27\pm 0.08$ & $1.33\pm 0.08$ \\
$\sigma^2$ & $0.08\pm 0.02$ & $0.09\pm 0.02$ & $0.05\pm 0.01$ &$0.06\pm 0.01$ \\
\hline
\end{tabular}
\caption{\label{tab2} Mean and variance of A$_{\rm LM}$, B$_{\rm LM}$ and $\sigma^2$ of the HIFLUGCS clusters for kT and NFW-F masses at $\Delta_c=200$ and $500$ without and with the baryon correction.}
\end{table*}

We compute the effective survey volume $V_{\rm eff}({M}_{\Delta})$ for each of the calibrated $\bar{L}_{\rm X}({M}_{\Delta})$ relations by evaluating Eq.~(\ref{vm_eff}) for the flat $\Lambda$CDM fiducial model of \citep{2017MNRAS.471.1370S} with $\Omega_M=0.27$ and $h=0.71$. We then compute from Eq.~(\ref{massfunction_def}) the corresponding halo mass functions assuming bins of size $\Delta\ln{M_{\Delta}}=0.3$.

\begin{figure}
    \includegraphics[width=1\columnwidth]{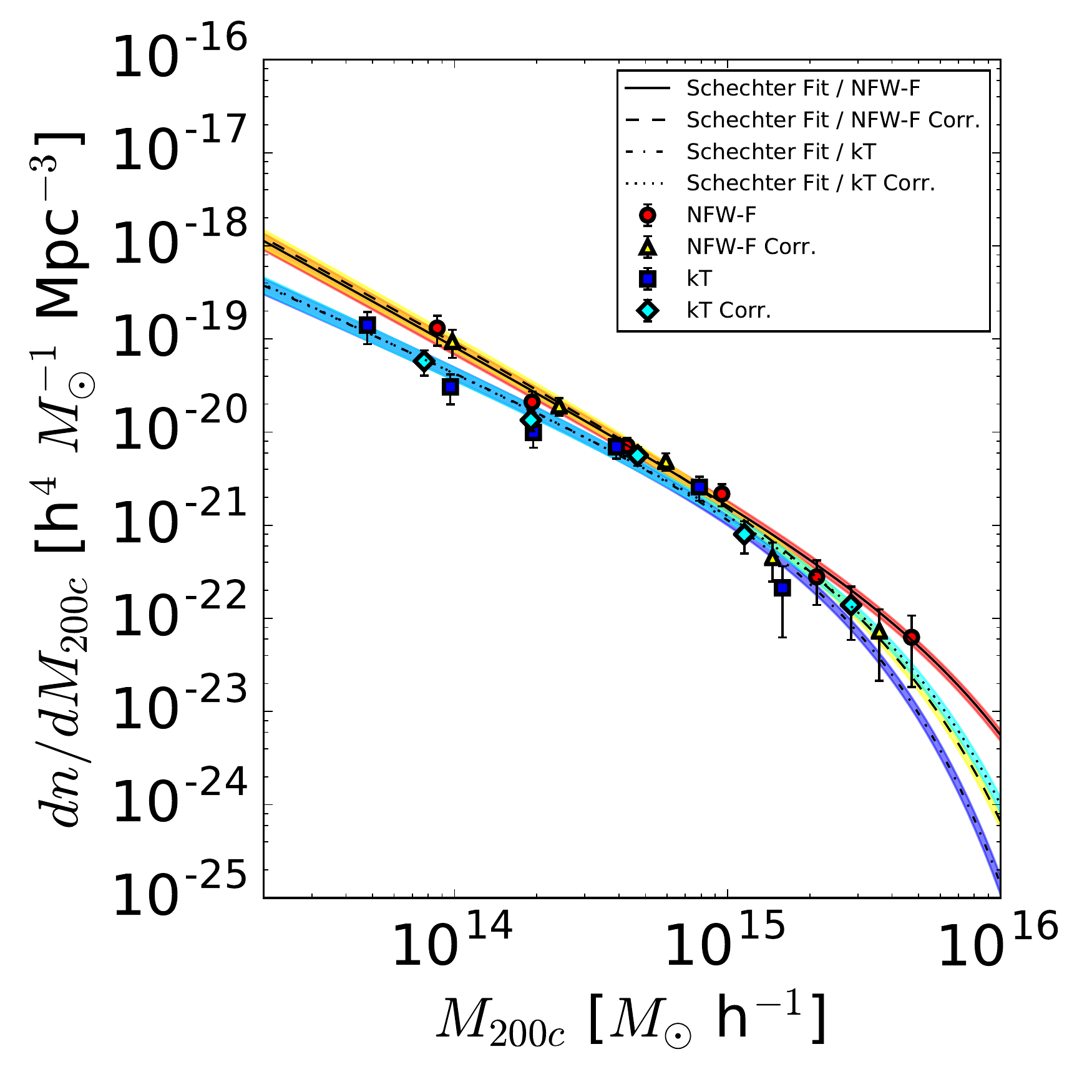}
    \includegraphics[width=1\columnwidth]{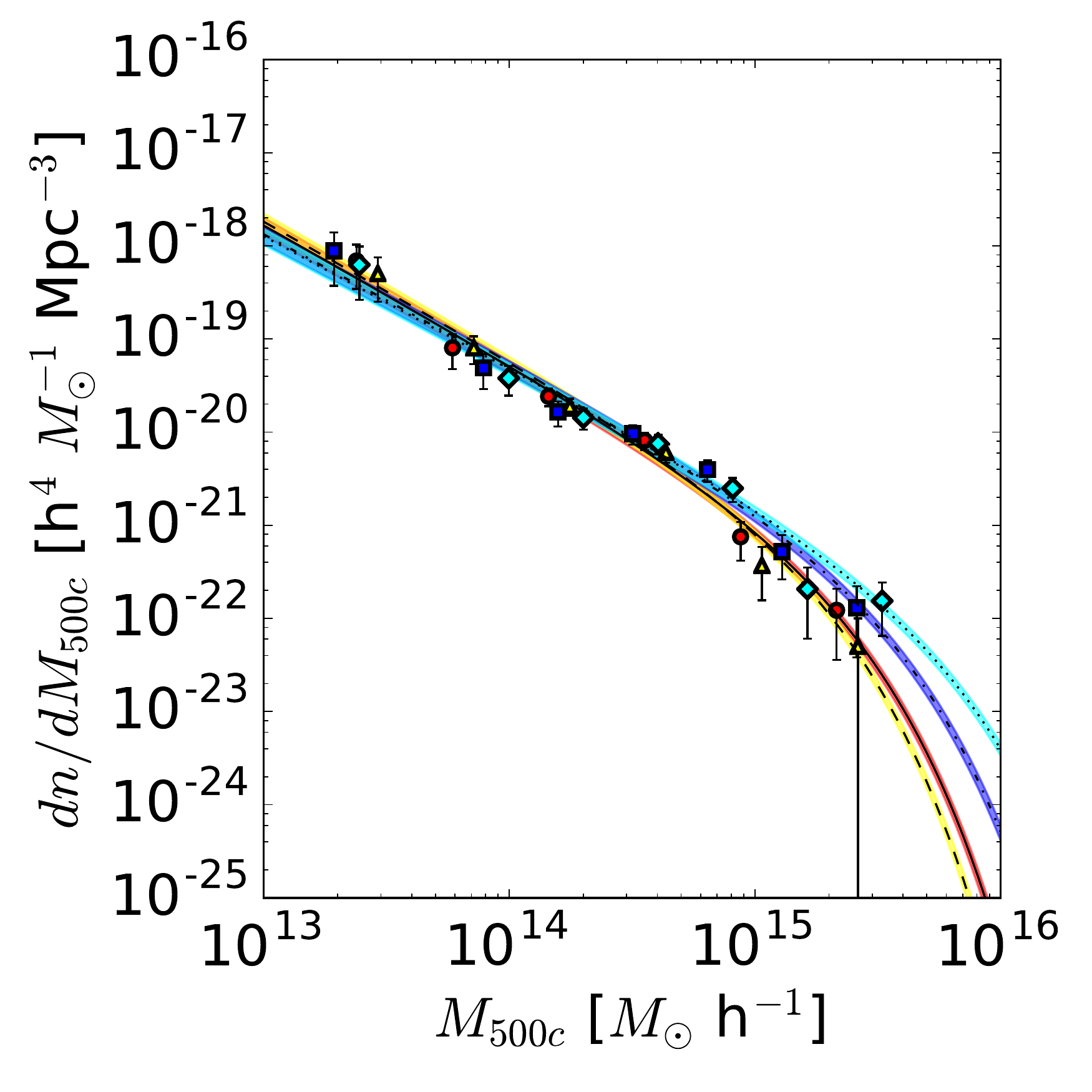}
    \caption{HIFLUGCS cluster mass function at ${\rm M}_{200c}$ (top panel) and ${\rm M}_{500c}$ (bottom panel) for the NFW-F masses without (red circles) and with baryon correction (yellow triangles), and the kT masses without (blue squares) and with (cyan diamonds) baryon corrections. The solid and dashed lines correspond to the best-fit Schechter-like functions, while the colored shaded areas delimit the $1\sigma$ uncertainty regions.}
    \label{fig5}
\end{figure}

In Fig.~\ref{fig5}, we plot the HIFLUGCS cluster mass functions at $\Delta_c=200$ (top panel) and $500$ (bottom panel) respectively. We fit the mass function data with a Schechter-like function given by:
\begin{equation}\label{schechter}
\frac{dn}{d\ln{M}_{\Delta}}=n_0\cdot {M}_{\Delta}\left(\frac{M_{\Delta}}{M_*}\right)^{\alpha}e^{-\frac{M_{\Delta}}{M_*}},
\end{equation}
where $n_0$, $\alpha$ and $M_*$ are coefficients which we determine using a Levenberg-Marquardt minimisation scheme. In Table~\ref{tab3} we quote the best fit values. We use these analytical fits to evaluate Eq.~(\ref{sparpred}). We have opted not to fit the data against the ST mass function used in Section~\ref{nbody}, rather to use the simple function given by Eq.~(\ref{schechter}) since this spares us from assuming a fiducial cosmology.
 
\begin{table*}
\centering
\begin{tabular}{c|c|c|c|c}
\hline
 & kT & kT Corr. & NFW-F & NFW-F Corr. \\
\hline
$\Delta_c=200$ & & & &\\
\hline
$n_{0}\,\left[10^{-21}\right]$ & $1.28390$ & $0.65743$ & $0.16647$ & $0.66796$  \\
$\alpha$ & $-1.3170$ & $-1.3538$ & $-1.6204$ & $-1.5971$ \\
$M_{*}\left[10^{15}\,M_{\odot}\,{\rm h}^{-1}\right]$& $1.5054$ & $2.2398$ & $4.6418$ &$2.2148$ \\
\hline
$\Delta_c=500$ & & & & \\
\hline
$n_{0}\,\left[10^{-20}\right]$ & $0.7913$ & $0.3288$ & $0.11801$ & $0.1793$ \\
$\alpha$& $-1.4063$ & $-1.4144$ & $-1.4869$ & $-1.4816$ \\
$M_{*}\left[10^{15}\,M_{\odot}\,{\rm h}^{-1}\right]$ & $1.9725$ & $3.4360$ & $1.3135$ & $1.0749$ \\
\hline
\end{tabular}
\caption{\label{tab3} Best-fit values of the Schechter-function coefficients for the HIFLUGCS mass functions at $\Delta_c=200$ and $500$ for the diffrent mass estimation schemes.}
\end{table*}

\subsection{HIFLUGCS Average Sparsity Relations}
We have computed the ensemble average sparsity of the HIFLUGCS clusters for the kT and NFW-F mass extrapolation schemes with and without baryon corrections. We have also computed the ratio of the averages of the inverse cluster masses Eq.~(\ref{sparsity_invmass}), the ratio of the averages of the cluster masses Eq.~(\ref{sparsity_averagemass}) and the average sparsity predicted by the mass function relation Eq.~(\ref{sparpred}). We quote the values of the different estimations in Table~\ref{tab_spars2}.

\begin{table*}
\centering
\begin{tabular}{|c|c|c|c|c|}
\hline
 & $\left\langle\frac{M_{200c}}{M_{500c}}\right\rangle$ & $\frac{\langle1/M_{500c}\rangle}{\langle1/M_{200c}\rangle}$ & $\frac{\langle M_{200c}\rangle}{\langle M_{500c}\rangle}$ & $\langle s^{\rm MF}_{200,500}\rangle$  \\
\hline
kT & $1.07\pm 0.06$ & $1.07\pm 0.02$ & $0.84\pm 0.04$ & $0.62\pm 0.05$ \\
kT Corr. & $1.07\pm 0.06$ & $1.03\pm 0.02$ & $0.86\pm 0.04$ & $0.71\pm 0.05$\\
NFW-F & $2.06\pm 0.06$ & $2.22\pm 0.06$ & $2.00\pm 0.13$ & $2.19\pm 0.03$  \\
NFW-F Corr. & $2.02\pm 0.06$ & $2.10\pm 0.07$ & $2.02\pm 0.13$ & $2.20\pm 0.03$ \\
\hline
\end{tabular}
\caption{\label{tab_spars2} Average sparsity estimates of the HIFLUGCS sample for different mass extrapolation schemes.}
\end{table*}

First, we may notice that the average sparsity values from the kT masses are about a factor of two smaller than the NFW-F case. As already mentioned, this is likely due to the extrapolation scheme. On the one hand, we have that the baryon corrected values only differ by a few percent compared to the uncorrected ones, which is a direct consequence of the fact that the cluster sample is dominated by massive clusters for which baryon correction are small. On the other hand, from the kT values quoted in \ref{tab_spars2} we can see that the different average sparsity estimates are inconsistent with each other at more than $2\sigma$. As such the masses inferred from the kT extrapolation scheme do not result in sparsity estimates whose ensemble average properties are coherent with expectations from N-body simulations.

In the NFW-F case, the consistency is within $2\sigma$. However, the NFW-F mass estimates implicitly assume the validity of NFW profile with a given concentration-mass relation. Hence, we would have expected a much better agreement among the average sparsity relations as in the case of the LoCuSS dataset presented in Section~\ref{locuss}. We believe that the remaining discrepancy may be due to residual systematics affecting the results of the extrapolation scheme at radii not covered by the X-ray observations or a consequence of the assumption of a specific concentration-mass relation in the estimation of the cluster masses. 

Overall, these results suggest that testing the average halo sparsity relations can provide a useful consistency check of the robustness of mass measurements of galaxy cluster samples. Along these lines it would be interesting to investigate whether different population of clusters affected by processes not included in $\Lambda$CDM N-body simulations may invalidate the consistency between the different average sparsity estimates. In such a case, the validity of these relations may be used as a criterion to identify outliers and select homogeneous cluster samples for cosmological data analyses. We leave such a study to future work.

\section{Conclusions}\label{conclu}
The dark matter halo sparsity is a direct observational proxy of the halo mass distribution. This characterizes halos in terms of the ratio of the masses enclosed within radii containing different overdensity thresholds. As shown in previous numerical studies a key feature of the halo sparsity is the fact that its ensemble average value can be predicted from prior knowledge of the halo mass function at the overdensities of interests. This relation has been derived as consequence of the fact that the halo sparsity does not significantly evolve with the halo mass. However, this is only one of the different ensemble average relations of the halo sparsity.

We have specifically focused on halo masses at $\Delta_c=200$ and $500$ for cosmological applications to galaxy cluster sparsity data analysis. Using halo catalogs from the MultiDark-Planck2 and Raygal simulations, we have shown that the ensemble average sparsity estimated from the geometric mean of individual halo sparsities in the ensemble coincides with the ratio of the averages of the inverse halo masses as well as the ratio of the averages of the halo masses at the overdensities of interest to within $1\%$ level. The validity of these different average sparsity estimates is a direct consequence of the fact that the halo sparsity $s_{200,500}$ is nearly uncorrelated with $M_{500c}^{-1}$ (which implies the consistency with the average sparsity value inferred from the mass function relation), as well as $M_{200c}$. 
The validity of these average sparsity relations suggests their use as a consistency check of the robustness of mass measurements in galaxy cluster samples. To this purpose we have tested these relations on the LoCuSS and HIFLUGCS cluster datasets respectively. The former consists of a sample of clusters with lensing mass estimates inferred from measurements of the shear profile of galaxy clusters. The latter consists of flux selected sample of local X-ray galaxy clusters with hydrostatic mass estimates at the overdensities of interests obtained with different extrapolation schemes.

We find that the average sparsity relations of the LoCuSS clusters are valid well within the uncertainties of the mass estimation errors. Moreover, we found the ensemble average value to be consistent with that expected from the N-body simulations.
In the HIFLUGCS case we have also accounted for the effects of baryons on the cluster mass estimates using results from hydrodynamics simulations. We have tested the average sparsity relation for two different extrapolation schemes used to infer the hydrostatic masses. In particular we consider masses obtained with the kT scheme, which relies on an analytical parametrization of temperature and surface density profiles, and masses inferred assuming a NFW density profile with an empirical median concentration-mass relation from numerical simulations. Independently of the baryon correction, we find that in the former case the ratio of the average halo masses and the average sparsity from the mass function relation are inconsistent with each other as well as with the ensemble average value from the geometric mean of the individual halo sparsities at more than $2\sigma$. Such inconsistency points to the extrapolation scheme as source of systematic errors due to the extrapolation of the cluster mass at large radii well beyond the radial interval covered by X-ray observations. The discrepancies among the different average sparsity estimates are less significant for the NFW-F masses, for which we find an agreement within $2\sigma$. This is not surprising since the NFW-F masses implicitly assume the NFW profile. However, given the relation between halo sparsity and concentration parameter for NFW halos, we would have expected a much better agreement. Hence, this may point to residual systematics affecting the extrapolation beyond the radial interval covered by the X-ray observations or an effect of the assumed median concentration-mass relation.

The work presented here suggests that the consistency of the average halo sparsity relations can be a powerful tool for testing the robustness of mass measurements in galaxy cluster samples (or alternatively point to deviations from the $\Lambda$CDM N-body simulation results).

\section*{Acknowledgements}
PSC is grateful to Stefano Ettori, Mauro Sereno, Gustavo Yepes, Thomas Reiprich and Stefano Andreon for useful discussions and suggestions. The authors are grateful to the anonymous referee for the useful comments, suggestions and constructive criticism that helped us to improve the work presented here. The research leading to these results has received funding from the European Research Council under the European Union's Seventh Framework Programme (FP/2007--2013) / ERC Grant Agreement n.~279954. We acknowledge support from the DIM-ACAV of the Region \^Ile-de-France. This work was granted access to the HPC resources of TGCC under allocation 2016-042287 made by GENCI (Grand \'Equipement National de Calcul Intensif). The CosmoSim database used in this paper is a service by the Leibniz-Institute for Astrophysics Potsdam (AIP). The MultiDark database was developed in cooperation with the Spanish MultiDark Consolider Project CSD2009-00064. The author gratefully acknowledge the Gauss Centre for Supercomputing e.V. (www.gauss-centre.eu) and the Partnership for Advanced Supercomputing in Europe (PRACE, www.prace-ri.eu) for funding the MultiDark simulation project by providing computing time on the GCS Supercomputer SuperMUC at Leibniz Supercomputing Centre (LRZ, www.lrz.de). Data visualisations are prepared with the \textsc{\mbox{matplotlib}}\footnote{\url{http://matplotlib.org/}} library \citep{Hunter07}, and DAG diagrams are drawn using the \textsc{dot2tex}\footnote{\url{https://dot2tex.readthedocs.org/}} tool created by K.~M.~Fauske and contributors.

\bibliographystyle{mnras}
\bibliography{refs}

\bsp	
\label{lastpage}
\end{document}